\documentclass[pre,aps,superscriptaddress,showpacs,floatfix,preprint]{revtex4-1}
\usepackage{amssymb,mathrsfs,amsfonts}
\usepackage{amsmath}
\usepackage{graphicx}
\usepackage{epstopdf}
\usepackage{nicefrac}
\usepackage{listings}
\usepackage{upgreek}
\usepackage{physics}
\usepackage{multirow}
\usepackage{color}
\usepackage{textcomp}
\usepackage{bibnames}
\usepackage{appendix}
\usepackage{epstopdf}
\usepackage{nicefrac}
\usepackage{listings}
\usepackage{upgreek}
\usepackage{physics}
\usepackage{multirow}
\usepackage{color}
\usepackage{natbib}
\usepackage{graphicx}
\usepackage{tabls}
\usepackage{afterpage}
\usepackage{amsthm}
\usepackage{lastpage}
\usepackage{empheq, etoolbox}
\usepackage{epstopdf}
\usepackage{changes} %
\colorlet{Changes@Color}{red}  

\newcommand{\A}{{\cal A}}
\newcommand{\B}{{\cal B}}

\newcommand{\vk}{{\varkappa_0}}
\newcommand\hF{\prescript{}{2}{\cal F}_1}
\newcommand{\RN}[1]{%
  \textup{\uppercase\expandafter{\romannumeral#1}}%
}

\newcommand{\hpsi}{\hat{\psi}_{s}}
\newcommand{\hsigma}{\hat{\sigma}^{s}}
\newcommand{\halpha}{\hat{\alpha}_{s}}
\newcommand{\hD}{\hat{D}_{s}}
\newcommand{\hkappa}{\hat{\varkappa}_{s,0}}
\newcommand{\hc}{\hat{c}_{s}}
\newcommand{\hDN}{\hat{D}_{s,N}}

\begin{document}
\title{The Time-Dependent Asymptotic $P_N$ Approximation\\for the Transport Equation} 
\author{Re'em Harel}
\affiliation{Department of Physics, Bar-Ilan University, 5290002, Ramat-Gan, Israel}
\author{Stanislav Burov}
\affiliation{Department of Physics, Bar-Ilan University, 5290002, Ramat-Gan, Israel}
\author{Shay I. Heizler}
\email{highzlers@walla.co.il}
\affiliation{Racah Institute of Physics, The Hebrew University, 9190401 Jerusalem, Israel}

\begin{abstract}
In this study a spatio-temporal approach for the solution of the time-dependent Boltzmann (transport) equation is derived. Finding the exact solution using the Boltzmann equation for the general case is generally an open problem and approximate methods are usually used. One of the most common methods is the spherical harmonics method (the $P_N$ approximation), when the exact transport equation is replaced with a closed set of equations for the moments of the density, with some closure assumption. Unfortunately, the classic $P_N$ closure yields poor results with low-order $N$ in highly anisotropic problems. Specifically, the tails of the particle's positional distribution as attained by the $P_N$ approximation, are inaccurate compared to the true behavior. In this work we present a derivation of a linear closure that even for low-order approximation yields a solution that is superior to the classical $P_N$ approximation. This closure, is based on an asymptotic derivation, both for space and time, of the exact Boltzmann equation in infinite homogeneous media. We test this approximation with respect to the one-dimensional benchmark of the full Green function in infinite 
media. The convergence of the proposed approximation is also faster when compared to (classic or modified) $P_N$ approximation.
\end{abstract}

\maketitle

\section{Introduction} 
\label{intro}
When modeling a thermo-dynamical system, that is out of equilibrium such as gas or fluid with temperature gradients, the most common equation used is the Boltzmann equation. This integro differential transport equation describes the local density of particles traveling and interacting inside a medium. It is frequently used for gas dynamics~\cite{Duderstadt}, charged particles transport inside plasma~\cite{epperlein1994}, radiative transport of photons in supernova~\cite{castor2004,pomraning2005equations}, inertial confinement fusion (ICF)~\cite{lindl2004,rosen1996,prr}, nuclear reactor physics~\cite{Duderstadt,CaseZweifel1967,BellGlasstone}, and persistent random walk~\cite{weiss2002,masoliver}. Specific limits of this equation give rise to other extremely popular frameworks of systems out of equilibrium, e.g. Fokker-Planck Equation~\cite{pawula1967approximation}. In most cases the solution of the time-dependent Boltzmann equation is challenging, and several approximation methods were devised over the years to cope with this problem. Unfortunately, when dealing with extreme observable, such as far tails of the positional distribution, a high-order approximation is essential for a proper description. And even then, the results are frequently poor when compared to the true behavior. In this manuscript, we address this problem by developing a different type of spatio-temporal approximation. 

In the mono-energetic homogeneous slab geometry, the Boltzmann equation attains the form:
\begin{equation} \label{transport_Eq}
    \frac{1}{v}\pdv{\psi(x, \mu, t)}{t} + \mu \pdv{\psi(x, \mu, t)}{x} + \sigma_t\psi(x, \mu, t) = \frac{1}{2}\int_{-1}^{1} \sigma_s(\mu_0) \psi(x, \mu', t) d\mu' + \frac{Q(x, t)}{2}
\end{equation}
Where $\mu$ is the cosine of the angle with respect to the $x$-axis ($\mu=\Omega \cdot \hat{x}$). $\psi(x, \mu, t)\equiv vn(x, \mu, t)$ is the angular flux in position $x$ in direction $\mu$ at time $t$. $n(x, \mu, t)$ is the local particle density and $v$ is the velocity of the particles. $\sigma_t=\sigma_s+\sigma_a$ is the total cross section which is composed of $\sigma_s(\mu_0)$ (where $\mu_0\equiv\Omega \cdot \Omega'$), the scattering cross section and $\sigma_a$, the absorbing cross section. $Q(x, t)$ is the source term.

There are several ways to solve the transport equation, among them are the spherical harmonics method (the $P_N$ approximation)~\cite{pomraning2005equations,BellGlasstone}, the discrete ordinates method (the $S_N$ method)~\cite{sn} and via Monte-Carlo simulations~\cite{imc}. The first two solve the deterministic transport equation, while the last one is statistical. In the $P_N$ approximation, the angular flux is constructed out of its first $N$ moments in the Legendre series (in one-dimension) or the complete spherical harmonics series (in multi-dimensions), yielding a set of moments equations for the exact angular flux. A modification of these methods is the main essence of this paper. All of the above methods are exact, for $N\to\infty$ in the deterministic cases, or when the number of histories (particles) goes to infinity, in the statistical one.

In order to reduce the complexity of solving the transport equation exactly, many approximations were proposed that handle the angular dependence.
As mentioned, one of the most well-known approximations is $P_N$ approximation, that can be derived by expanding the angular flux in a full set of spherical harmonics thus, transforming the transport equation to a set of infinite coupled equations, which is exact. To reduce the infinite set of equations to a finite set, a closure or a cutoff method is needed. For example, the classical $P_N$ approximation is obtained by introducing a closure that sets to zero all the coefficients of the expansion with index larger than $N$~\cite{Duderstadt,CaseZweifel1967,pomraning2005equations}. For weakly isotropic problems, low order of $N$ is sufficient. However, to describe highly anisotropic particles' distribution the classical $P_N$ approximation needs a relatively large number of $N$, which makes it hard to obtain especially in multi-dimensions. We note that advanced different closures may be chosen~\cite{levermore2011,mcclarren2016}. However, in most of them the implementation is complicated, due to the presence of explicit derivatives of the moments, and the non-linearity of the closures.

By assuming that the angular dependence is isotropic or nearly isotropic one can derive one of the most well-known approximation namely the {\em classical diffusion approximation} (the Eddington approximation)~\cite{Duderstadt,CaseZweifel1967,pomraning2005equations}.
The diffusion approximation satisfies the central-limit theorem (CLT)~\cite{weiss2002}.
In an alternative manner, the classic diffusion approximation corresponds to the most simple expression of the classical $P_N$ approximation, using $N=1$. However, the diffusion approximation fails to describe the particles' density in highly anisotropic scenarios. A well-known modification for this approximation is the {\em asymptotic diffusion approximation}~\cite{case1953introduction}. This approximation is based on an asymptotic spatial behavior and it successfully solves time-independent problems, such as the correct critical radii of bare reactor systems~\cite{BellGlasstone}, using the correct asymptotic (infinite-media) transport eigenvalues. Nevertheless, in time-dependent problems, both the classic and asymptotic diffusion approximations fail to describe the particles' front density, due to the parabolic nature of the diffusion equation, which predicts an infinite particle velocity. A full time-dependent $P_1$ approximation which tends to the telegrapher's equation, yields a finite velocity, but a wrong one, which eventually yields even worse results than diffusion~\cite{bengston1958}.

A related approximation was proposed by Heizler~\cite{heizler2010asymptotic}, namely the {\em asymptotic $P_1$ approximation} that is based on an asymptotic behavior in both space and time of the transport equation. This approximation, in some sense, a time-dependent equivalent to the asymptotic diffusion approximation. This approximation has a $P_1$ form that overcomes the parabolic nature of the asymptotic diffusion. It yields the correct time-independent eigenvalue of the exact transport equation, and produces velocity that is very close to the correct one. This approximation was derived and tested also for radiative transfer~\cite{heizler2012}, and was found to share similar asymptotic features as the $P_2$ approximation~\cite{heizler2012sp}. A modified version of this approximation, suitable for heterogeneous media was proposed by Cohen et al.~\cite{cohen2018discontinuous} namely the {\em discontinuous asymptotic $P_1$ approximation}. This approximation yields extremely good results in both radiative transfer benchmarks and Marshak-wave experiments~\cite{cohen2018discontinuous,cohen2019discontinuous}. 

However, there are still, non-negligible deviations between the asymptotic $P_1$ approximation and the exact transport solutions in homogeneous media, especially for anisotropic scenarios, specifically for the `tails' of the particles' distributions~\cite{heizler2010asymptotic}. These deviations are due to the low order of $N=1$ in this approximation. Therefore, a higher order of $N$ is needed for a more accurate modeling. In this paper the asymptotic $P_1$ methodology is generalized for any given finite $N$, namely, the time-dependent {\em asymptotic $P_N$ approximation}. The new approximation yields a good approximate behavior even for low-order $N$'s. Our work is a generalization of ideas introduced in the novel work of Pomraning~\cite{pomraning1964generalized}, that derived a time-independent asymptotic $P_N$ approximation as a general-$N$ expansion of the asymptotic diffusion approximation.  

At first, we introduce in Sec.~\ref{section_pn} the different approximations and variations of the $P_N$'s. Afterwards in Sec.~\ref{td_ap_n} we generalize the asymptotic $P_1$ methodology for any given $N$, deriving the time-dependent asymptotic $P_N$ approximation, and in Sec.~\ref{test_benchmark} we show that it converges much faster than the classic $P_N$ approximation for any given $N$ in a time-dependent benchmark, of the infinite media full Green function~\cite{ganapol1999,olson2004numerical}. We show that the new approximation yields better results than other modified approximations and closures for the $P_N$ equations.

\section{Different $P_N$ Approximations} \label{section_pn}
In this section we present different $P_N$ approximations, that are exploited in this study, including the advanced steps that had been done to modify the classic closure of the $P_N$ approximation. Specifically, we present the time-dependent asymptotic $P_1$ of Heizler~\cite{heizler2010asymptotic} and the time-independent asymptotic $P_N$ of Pomraning~\cite{pomraning1964generalized}, the two foundations that the new approximation is based on. 

\subsection{The Classic Time-Dependent and Independent $P_N$ Approximation}
As mentioned in the introduction, the $P_N$ approximation can be derived by expanding the angular flux in a full set of spherical harmonics~\cite{Duderstadt,CaseZweifel1967,pomraning2005equations}, or in one-dimension in a full set of Legendre polynomials:
\begin{equation}\label{polynomseries}
    \psi(x, \mu, t) = \sum_{n=0}^{\infty}\frac{2n+1}{4\pi} \psi_n(x, t) P_n(\mu),
\end{equation}
where $P_n(\mu)$ are the Legendre polynomials and the angular moments are given by:
\begin{equation}
    \psi_n(x,t) = 2\pi\int_{-1}^{1} P_n(\mu) \psi(x, \mu, t)d\mu
\end{equation}

The first two moments are $\psi_0(x, t)\equiv\phi(x, t)$ - the scalar flux and $\psi_1(x, t)\equiv J(x, t)$ - the particles' current. Substituting Eq.~\ref{polynomseries} back into the transport equation (Eq.~\ref{transport_Eq}) and using the orthogonal property of Legendre polynomials yields an infinite coupled set of equations:
\begin{equation}\label{pnequations}
\frac{1}{v} \pdv{\psi_n(x, t)}{t} + (\sigma_t-\sigma_s^{(n)}) \psi_n(x, t) +\left(\frac{n+1}{2n+1}\right)  \pdv{\psi_{n+1}(x, t)}{x} + \left(\frac{n}{2n+1}\right) \pdv{\psi_{n-1}(x, t)}{x} = Q^{(n)}(x, t)
\end{equation}
Where $\sigma_s^{(n)}$ is the $n$'th moment of the scattering cross-section and $Q^{(n)}(x, t)$ is the $n$'th moment of the source term. Both of these terms can be derived in the same way that $\psi_n(x,t)$ was derived, i.e by expansion to Legendre series. For the {\em time-independent} case, the first term in Eq.~\ref{pnequations} is dropped, and so is the time-dependency of the different moments.

In order to reduce the infinite coupled set of equations and obtain a finite number of equations with the same number of variables, some sort of closure needs to be introduced to the equations. The classic $P_N$ approximation closure is obtained by setting:
\begin{equation}
    \pdv{\psi_{n+1}(x, t)}{x}=0, \qquad\qquad n\geqslant N
\end{equation}
By introducing this classical closure, we obtain $N-1$ exact equations and one equation that is an approximated one. This closure transforms the $N$'th formula in Eq.~\ref{pnequations} to:
\begin{equation}\label{pn_classic_closure}
    \frac{1}{v}\pdv{\psi_N(x,t)}{t} + (\sigma_t - \sigma_s^{(N)})\psi_N(x,t) + \left(\frac{N}{2N+1}\right)\pdv{\psi_{N-1}}{x} = Q^{(N)}(x,t)
\end{equation}
While keeping all other formulas with $n<N$ unchanged.
We now describe several concrete examples.

The derivation from here on will use the notation of isotropic scattering term, i.e. the total cross-section $\sigma_t$. However, we note that including the general case for anisotropic scattering cross-section, can be done by replacing $\sigma_t$ with $\sigma_t - \sigma_s^{(n)}$. Also, the derivation is presented for isotropic sources only, i.e., $Q(x,t)=Q^{(0)}(x,t)$. The derivation here requires that the closure equation will be source-free, so the $N$'th moment of the source must be neglected.

\subsubsection{$P_1$}
The most simple example of the classic $P_N$ approximation is the $P_1$ approximation. The $P_1$ approximation can be obtained by following the above procedure and setting $N=1$ yielding two equations with two unknown variables. The first equation, which is an exact equation and the second equation is an approximated equation, are given by:
\begin{subequations}
    \begin{equation} \label{p1firstequation}
        \frac{1}{v} \pdv{\phi(x, t)}{t} + \pdv{J(x, t)}{x}+ \sigma_a \phi(x, t) = Q(x, t)
    \end{equation}
    \begin{equation} \label{p1secondequation}
        \frac{1}{v}\pdv{J(x,t)}{t} + \frac{1}{3}\pdv{\phi}{x} + \sigma_t J(x,t) = 0
    \end{equation}
\end{subequations}
As mentioned in Sec.~\ref{intro} the $P_1$ approximation, in a homogeneous medium, tends to the Telegrapher's equation, and has a hyperbolic nature, which means that it yields a finite particle velocity~\cite{heizler2010asymptotic}. The $P_1$ approximation results in a particle velocity that is equal to $v/\sqrt{3}$, that unfortunately fails to describe the correct transport behavior~\cite{bengston1958}.

\subsubsection{Diffusion}
The classical diffusion approximation can be seen as a special case of the $P_1$ approximation and it can be obtained by neglecting the time derivative of the particles' current ($J(x, t)$) in Eq.~\ref{p1secondequation}:
\begin{equation}
    \frac{1}{\vert J(x,t)\vert}\pdv{\vert J(x,t)\vert}{t} \ll v\sigma_t
\end{equation}
Eq.~\ref{p1secondequation} takes a Fick's law form:
\begin{equation} 
    J(x,t) = -D(x)\pdv{\phi}{x}
\end{equation}
With the well known diffusion coefficient $D(x) = 1/3\sigma_t$. By substituting Eq.~\ref{p1secondequation} into the exact Eq.~\ref{p1firstequation} we derive time-dependent diffusion equation:
\begin{equation}\label{diffusion}
    \frac{1}{v} \pdv{\phi(x, t)}{t}- \pdv{}{x}\left(D(x)\pdv{\phi(x, t)}{x}\right)+ \sigma_a \phi(x, t)  = Q(x, t)
\end{equation}
The natural solution for the Green function takes a Gaussian shape (i.e. normal distribution) thus, yields an infinite particle velocity, as a result of the parabolic nature of the diffusion equation.

\subsubsection{$P_2$}
The second order of the $P_N$ approximation is obtained for $N=2$, yielding the following two exact equation and a third that is approximated:
\begin{subequations}
    \begin{equation} \label{p2a}
        \frac{1}{v} \pdv{\phi(x, t)}{t} + \pdv{J(x, t)}{x}+ \sigma_a \phi(x, t) = Q(x, t)
    \end{equation}
    \begin{equation} \label{p2b}
        \frac{1}{v}\pdv{J(x,t)}{t} + \sigma_t J(x,t) + \frac{1}{3}\pdv{\phi}{x} + \frac{2}{3}\pdv{\psi_2}{x} = 0
    \end{equation}
    \begin{equation} \label{p2c}
         \frac{1}{v}\pdv{\psi_2(x,t)}{t} + \sigma_t \psi_2(x,t) + \frac{2}{5}\pdv{J}{x} = 0
    \end{equation}
\end{subequations}
However, Davison~\cite{davison1957neutron} and Pomraning~\cite{pomraning2005equations} claim that in general, even-order $P_N$ approximations have difficulties that the odd-order do not have, and that $P_N$ with odd $N$ is more accurate than $P_{N+1}$. Since each $P_N$ approximation is equivalent to $S_{N+1}$ approximation, even order $P_{N-1}$ yields a perpendicular direction ($\mu_n=0$) (due to the odd order $S_N$), thus the streaming term is disappearing, so the equation is no longer differential in space (and, as such, cannot be involved in the boundary conditions). Thus, the even-order $P_N$ approximation is rarely used in practice. Nevertheless, Ravetto and others~\cite{CoppaRavetto1980,RulkoLarsen1993,heizler2012sp,ShinMorel1993,PomraningRulkoSu1994,Rulko1995}, argue that there are benefits in using an even order of $N$, specifically the $P_2$ approximation. In this work we will also present a modified asymptotic $P_2$ approximation that yields excellent results.

\subsection{Dawson (navy) Modified $P_2$ Approximation}\label{dawson_section}
Several modifications to the classical $P_N$ scheme were proposed over the years. Of special interest is the modification to the classic $P_2$ approximation that was suggested by Dawson in a DTMB (navy) report~\cite{dawson} (for the time-independent case only). In his work, Dawson proposes two coefficients that multiply the $\sigma_t$ terms in the time-independent version of Eqs.~\ref{p2b} and~\ref{p2c}. A time dependent modification of Dawson's approximation is:
\begin{subequations}
\label{dawson_sp2_eq}
   \begin{equation} 
        \frac{1}{v} \pdv{\phi(x, t)}{t} + \pdv{J(x, t)}{x}+ \sigma_a \phi(x, t) = Q(x, t)
    \end{equation}
   \begin{equation}
        \frac{1}{v}\pdv{J(x,t)}{t} + \alpha_1 \sigma_t J(x,t) + \frac{1}{3}\pdv{\phi}{x} + \frac{2}{3}\pdv{\psi_2}{x} = 0
    \end{equation}
    \begin{equation}
         \frac{1}{v}\pdv{\psi_2(x,t)}{t} + \alpha_2 \sigma_t \psi_2(x,t) + \frac{2}{5}\pdv{J}{x} = 0
    \end{equation}
\end{subequations}
This approximation is based on the discrete ordinates-like method ($S_3$-like approximation), choosing the cosine of the discrete directions and the weights (for the quadrature formula that calculates the moments). The values of $\alpha_{1,2}$ is determined such that it reproduces several physical quantities (under additional assumptions that are specified in~\cite{dawson}). Dawson's proposed two different values for determining $\alpha_{1,2}$:
\begin{itemize}
\item{$\alpha_1 = 1.18858$ and $\alpha_2=4.16349$. This choice is based on choosing the quadrature directions and weights that minimize the error to transport kernels of the integral form of the transport equation. This choice will be marked in section~\ref{test_benchmark} as Navy $P_2$ a, and marked as the favorite choice by Dawson.}
\item {$\alpha_1 = 1$ and $\alpha_2=2.4$. This choice is based on choosing the double Gauss quadrature set (for yielding the integral moments correctly). This choice will be marked in section~\ref{test_benchmark} as Navy $P_2$ b.} 
\end{itemize}
For a detailed discussion of how to choose the coefficients $\alpha_{1,2}$, see~\cite{dawson}. In Dawson's time-independent tests, both of these different values yield better results than both $P_1$ and $P_2$ compared to the exact solution, which lies between the $P_1$ and the $P_2$ results.

\subsection{Pomraning's Time-independent Asymptotic $P_N$}
Since the classical $P_N$ approximation, will accurately describe the behavior in anisotropic particles' distribution given $N$ is large enough, Pomraning proposed a new general closure method that replaces the classic $P_N$ closure~\cite{pomraning1964generalized}. The new closure method retains the symmetry of the transport equation~\cite{pomraning2005equations}. In this approximation, the last approximated equation takes the following form:
\begin{equation}\label{pomraningtruncation}
        \sigma_t\psi_N(x) + \left(\frac{N}{2N+1}\right)\pdv{\psi_{N-1}(x)}{x} + \left(\frac{N + 1}{2N + 1}\right) \pdv{(\alpha_N \psi_{N-1}(x))}{x} =0
\end{equation}
$\alpha_N$ is defined as:
\begin{equation}\label{alphapomraning0}
 \alpha_N = \frac{\int_{-1}^{1} \psi_a(x,\mu) P_{N+1}(\mu) d\mu}{\int_{-1}^{1} \psi_a(x,\mu) P_{N-1}(\mu) d\mu}
\end{equation}
$\psi_a$ is the assumed form of the directional flux. By setting $\alpha_N$ to zero, we obtain the classical $P_N$ approximation.
When $\psi_a$ is the exact solution of the Boltzmann equation, Eqs.~\ref{pomraningtruncation} and ~\ref{alphapomraning0} are exact, and this finite $N$ approximation yields the exact solution.

While $\psi_a$ is not known in the general case, Pomraning~\cite{pomraning1964generalized} introduced several different ways to obtain $\psi_a$, each one is suitable for a different (angular distribution) case. The most accurate expression for $\psi_a$ (for general purposes) is the asymptotic distribution in infinite homogeneous media, which can be derived by solving the time-independent (source-free) Boltzmann equation for infinite homogeneous medium:
\begin{equation}\label{boltzmman_indep}
    \sigma_t\psi(x,\mu) + \mu \pdv{\psi(x,\mu)}{x} = \frac{c\sigma_t}{2}\int_{-1}^{1} \psi(x, \mu', t) d\mu'
\end{equation}
Where $c=\sigma_s^{(0)}/\sigma_t$ is the mean number of particles that are emitted per collision, and is called the {\em albedo}. By separation of variables, one can solve this equation and obtain a closed transcendental equation for $\varkappa_0$ - the eigenvalues of the asymptotic solution in $x$~\cite{case1953introduction,CaseZweifel1967}:
\begin{equation}
    \frac{2}{c}=\frac{1}{\varkappa_0}\ell n \left(\frac{1+\varkappa_0}{1-\varkappa_0}
    \right)
\label{kappa}
\end{equation}
By normalizing the asymptotic expression for the angular eigenfunctions - $\psi_a$ yields: 
\begin{equation}\label{asymptotc_eq}
\psi(x,\mu)\approx\psi_a(x,\mu)=\frac{c}{2}A_0\frac{e^{\varkappa_0\sigma_tx}}{1+\mu\varkappa_0}+
\frac{c}{2}B_0\frac{e^{-\varkappa_0\sigma_tx}}{1-\mu\varkappa_0}
\end{equation}
where $A_0$ and $B_0$ are determined from the boundary conditions.

By substituting the asymptotic flux (Eq.~\ref{asymptotc_eq}) into the closure term (Eq.~\ref{alphapomraning0}), yields:
\begin{equation}\label{alphapomraning}
\alpha_N = \frac{\int_{-1}^{1} \left(\frac{1}{1+\mu\varkappa}\right) P_{N+1}(\mu) d\mu}{\int_{-1}^{1} \left(\frac{1}{1+\mu\varkappa}\right) P_{N-1}(\mu) d\mu}
\end{equation}
Substituting Eq.~\ref{alphapomraning} back to the closure equation (Eq.~\ref{pomraningtruncation}), yields the time-independent asymptotic $P_N$ approximation. By definition, the largest eigenvalue of the asymptotic $P_N$ equals to the asymptotic eigenvalue of the exact time-independent Boltzmann equation~\cite{pomraning1964generalized}. While in Ref.~\cite{pomraning1964generalized}, the analysis was introduced for slab geometry, in Ref.~\cite{pomraning1965asymptotically} Pomraning expands his analysis to multi-dimensional case with general geometry.
This approximation is a generalization of the asymptotic diffusion for a general $N$. For $N=1$, this approximation reproduces the asymptotic diffusion approximation.

\subsubsection*{Asymptotic Diffusion}
The Pomraning time-independent asymptotic $P_N$ approximation for the $N=1$ case, takes the following form:
\begin{subequations}
\begin{equation}
    \pdv{\psi_1}{x} + (1-c)\psi_0 = Q(x) 
\end{equation}
\begin{equation}
    \psi_1 = -\pdv{}{x}\left( \frac{1-c}{\varkappa_0^2}\psi_0\right).
\label{AD_Fick}
\end{equation}
\end{subequations}
This of course gives rise to the asymptotic diffusion approximation with the well-know asymptotic diffusion coefficient:
\begin{equation}\label{D0}
 D(c) = \frac{1-c}{\sigma_t \varkappa_0^2(c)} \equiv \frac{D_0(c)}{\sigma_t}
\end{equation}
$D_0(c)$ is a dimensionless diffusion coefficient which is a function of the asymptotic eigenvalue $\varkappa_0$(c) (Eq.~\ref{kappa}), and has several fits for different range-values of $c$. The fits that were used in the this work are shown in the appendix.

As one can see, the asymptotic diffusion can be seen as special case of Pomraning's time-independent asymptotic approximation with $N=1$. As a matter of fact, Eq.~\ref{AD_Fick} reproduces the {\em new discontinuous diffusion theory} of Pomraning~\cite{pomraning2005equations}, where the diffusion coefficient is inside the space derivative, instead of the classic notation of the Fick's law, standard or asymptotic, which has following form:
\begin{equation}
    \psi_1 = -\left( \frac{1-c}{\varkappa_0^2}\right)\pdv{\psi_0}{x}\equiv -D(c)\pdv{\psi_0}{x}
\label{AD_Fick1}
\end{equation}

The asymptotic diffusion approximation has a complementary asymptotic derivation of the boundary conditions~\cite{case1953introduction,winslow1968}, allowing solving finite problems, and may be used in time-independent problems, such as the bare critical radius of a material~\cite{BellGlasstone}.
However, it can also be used in time-dependent problems~\cite{winslow1968} (although, it is a parabolic equation that yields infinite particle velocity). The generalization to multi-group is trivial, using a group-dependent $c_g$:
\begin{equation}
c_g=\frac{\sum_{g'}^G\sigma_{g'\to g}\phi_{g'}+Q_g}{\sigma_g\phi_g}
\label{group_albedo}
\end{equation}
Also, a discontinuous version of the asymptotic diffusion approximation was derived, in order to achieve better accuracy for multi-regions problems~\cite{zimmerman1979,PomraningRulkoSu1994,GanapolPomraning1996} (note that Pomraning's Fick's law, Eq.~\ref{AD_Fick}, the new diffusion theory~\cite{pomraning2005equations}, yields a discontinuous form also, due to the discontinuity in the asymptotic diffusion coefficient at the boundary between two different media).

Back to the asymptotic $P_N$, noticing the difference between Eqs.~\ref{AD_Fick} and~\ref{AD_Fick1}, we adopt Pomraning's closure (Eq.~\ref{pomraningtruncation}) with a slight modification, taking $\alpha_N$ outside of the spatial derivative (in order to follow the common asymptotic diffusion rationale, in the case of $N=1$):
\begin{equation}\label{pomraningliketruncation}
        \sigma_t\psi_N(x) + \left(\frac{N}{2N+1}\right)\pdv{\psi_{N-1}(x)}{x} + \left(\frac{N + 1}{2N + 1}\right)\alpha_N \pdv{\psi_{N-1}(x)}{x} =0
\end{equation}
In this case, forcing $N=1$, will yield the well-known notation of the asymptotic diffusion Fick's law, Eq.~\ref{AD_Fick1}, exactly.

The time-independent asymptotic $P_N$ approximation is not common in the literature. Huang et al.~\cite{huang1972asymptotic} have shown that the asymptotic $P_N$ preserves the asymptotic solution parameters - the eigenvalues, magnitudes and root-mean-square distances. Morel et al.~\cite{morel2013asymptotic} have found the directions and weights for an equivalent-$S_N$ calculation that yields the asymptotic solution.

\subsection{Heizler's Asymptotic $P_1$}
Another time-dependent $P_1$ approximation version was developed by one of us, Heizler~\cite{heizler2010asymptotic}, which we will later generalize to the time-dependent asymptotic $P_N$ approximation. The rationale behind Heizler's work is to find a modified $P_1$ closure based on the asymptotic solution of the time-dependent mono-energetic Boltzmann equation (Eq.~\ref{transport_Eq}), both in space and time. This approximation is the time-dependent version of the asymptotic diffusion in the time-independent case.

Applying Laplace transformation to Eq.~\ref{transport_Eq} yields:
\begin{equation}\label{s_boltzmann}
    \left(\frac{s}{v} + \sigma_t\right)\hpsi(x,\mu) + \mu \pdv{\hpsi(x,\mu)}{x} = \frac{c\sigma_t}{2}\int_{-1}^{1} \hpsi(x, \mu', t) d\mu'
\end{equation}
Defining the following `new' coefficients:
\begin{subequations}\label{Hnotation}
    \begin{equation}
        \hsigma_t = \sigma_t + \frac{s}{v} 
    \end{equation}
    \begin{equation}
        \hc = \frac{\sigma_s}{\hsigma_t} = \frac{c}{1 + \frac{s}{v\sigma_t}}    
    \end{equation}
\end{subequations}
and substituting the new coefficients back to Eq.~\ref{s_boltzmann} yields:
\begin{equation}
     \hsigma_t\hpsi(x,\mu) + \mu \pdv{\hpsi(x,\mu)}{x} = \frac{\hc\hsigma_t}{2}\int_{-1}^{1} \psi(x, \mu', t) d\mu'
\end{equation}
As one can notice, this is the exact same equation as Eq.~\ref{boltzmman_indep} with the `new' coefficients in the Laplace domain. Therefore, the asymptotic solution ($s$-dependent) is:
\begin{equation}\label{TD_asymptotc_eq}
\hpsi(x,\mu)\approx\frac{\hc}{2}A_0\frac{e^{\hkappa\hsigma_tx}}{1+\mu\hkappa}+\frac{\hc}{2}B_0\frac{e^{-\hkappa\hsigma_tx}}{1-\mu\hkappa},
\end{equation}
with $s$-dependent closed transcendental equation for the eigenvalues:
\begin{equation}
\frac{2}{\hc}=\frac{1}{\hkappa}\ell n\left(\frac{1+\hkappa}{1-\hkappa}\right)
\label{hkappa}
\end{equation}

With further analysis, a modified Fick's law form can be derived that is similar to the time-independent asymptotic $P_1$:
\begin{equation}\label{heizler_p1}
    {\hpsi{}_{,1}}(x) = - \frac{1-\hc}{\hsigma_t\hkappa^2} \pdv{\hpsi{}_{,0}}{x}, 
\end{equation}
with the diffusion coefficient:
\begin{equation}\label{heizler_d}
    \hat{D_s}(\hc,\hsigma_t) = \frac{1-\hc}{\hkappa^2 \hsigma_t} = v\frac{v\sigma_t(1-c) + s}{(v\sigma_t + s)^2 \hkappa^2} \equiv v\frac{\hat{D}_0(\hc)}{v\sigma_t+s}
\end{equation}
$\hat{D}_0(\hc)$ is the same diffusion coefficient as provided by Eqs.~\ref{D0_small_c} and \ref{D0_big_c} but it is a function of the $s$-dependent albedo $\hc$.

The work of Heizler proposed a new form of the last approximated equation, that includes two media-dependent variables $\A(c)$ and $\B(c)$, of the following form~\cite{heizler2010asymptotic}:
\begin{equation}\label{shay_secondAB}
    \frac{\A(c)}{v} \pdv{\psi_1(x,t)}{t} + \pdv{\psi_0}{x} + \B(c) \sigma_t \psi_1(x,t) = 0
\end{equation}
Applying Laplace transform to Eq.~\ref{shay_secondAB} yields an $s$-dependent Fick's law that is a function of $\A(c)$ and $\B(c)$:
\begin{equation}\label{s_ficks}
    \hat{D}_s(\A,\B) \equiv \frac{v}{\A(c) s + \B(c) v\sigma_t}
\end{equation}
By introducing the approximate Eq.~\ref{s_ficks} into the exact Eq.~\ref{heizler_d}, and expanding $\hat{D}_0(\hc)$ in a Taylor series for small $s$ (asymptotic in time), one can solve for $\A(c)$ and $\B(c)$. $\B(c)\equiv1/D_0(c)$ of the asymptotic diffusion theory, and produces the correct time-independent eigenvalues, while $\A(c)$ is responsible for the correct temporal behavior. For example, for purely scattering medium ($c=1$), the classic $P_1$ forces $\A=\B=3$, while the asymptotic $P_1$ yields $\B=3$ and $\A=3/5$. Note that the {\em ad hoc} $P_{1/3}$ approximation~\cite{p13} forces $\B=3$ and $\A=1$ which is quite close to the asymptotic $P_1$ approximation.

In Ref.~\cite{heizler2010asymptotic} it can be seen that in the full Green function source problem ($Q(x,t) = Q_0 \delta(x)\delta(t)$) in an infinite homogeneous medium, the asymptotic $P_1$ yields good accuracy, compared to the exact solution, better than the asymptotic diffusion approximation (which yields an infinite particle velocity), and much better than the classic $P_1$ approximation. The asymptotic $P_1$ approximation was tested also in a local thermodynamic equilibrium (LTE) radiative transfer problems, yielding relatively good results, especially in vicinity of the tails~\cite{heizler2012}.
The asymptotic $P_1$ approximation was also found to share the same asymptotic behavior as the $P_2$~\cite{heizler2012sp} approximation. Indeed, in the Green function problem, $P_2$ yields much better results than the classic $P_1$, and close to the asymptotic $P_1$ results.

Additional major advance was the derivation of a {\em discontinuous asymptotic $P_1$ approximation}~\cite{cohen2018discontinuous}. In this version, a modified discontinuous $P_1$ closure is offered, extending the validity of the asymptotic $P_1$ to heterogeneous media. This approximation yields a discontinuity in a sharp boundary between two media, yielding the correct asymptotic solution on each side. This is extremely important in radiative transfer problems, where the cross section is temperature dependent. This approximation yields better results than all other approximations, even better than the gradient-dependent approximations, such as the Flux-Limiters or the Variable Eddington Factors approximations~\cite{cohen2018discontinuous,cohen2019discontinuous}.

Still, even in homogeneous medium, there is still a non-negligible deviation between the asymptotic $P_1$ approximation and the exact solution~\cite{heizler2010asymptotic}. Thus, a general $N$ extension is required, for obtaining a more accurate results, for sufficiently low $N$. This is what motivates the derivation of the time-dependent asymptotic $P_N$ considered in the next section.

\section{Time-dependent Asymptotic $P_N$}\label{td_ap_n}
In this section we derive the generalization for Heizler's asymptotic $P_1$ approximation~\cite{heizler2010asymptotic}, to a general $N$ namely, the time-dependent asymptotic $P_N$ approximation. The derivation will be the time-dependent equivalent to Pomraning's time-independent derivation~\cite{pomraning1964generalized}.

We note that there are several alternate closures for the $P_N$ equations that multiplies either the time or spatial derivative by linear scale factors, for a faster convergence of the $P_N$~\cite{olson2012alternate, olson2019positivity}. For example, the scale factors can be chosen such that waves propagate at exactly the speed of the particles'. However, these approximations are purely {\em ad hoc} approximations, which are scaled each time to a specific problem. Advanced closures may also be used~\cite{levermore2011,mcclarren2016}, however, the implementation of most of them is much more complicated, due to the non-linearity of the closure equations. 

\subsection{Deriving the Time-dependent Asymptotic $P_N$}
\label{heart}

The general way to obtain the time-dependent asymptotic $P_N$ is done by modifying the last approximated equation, as the $n < N$ equations are the exact $P_n$ equations.
Firstly, we introduce a time-dependent closure equation (to the $N$-equation), which is similar to Pomraning's time-independent closure asymptotic with a similar modification to the one presented in Eq.~\ref{pomraningliketruncation}, i.e, taking the $\alpha_N$ outside of the spatial derivative:
\begin{equation}\label{atd_nequation}
        \frac{1}{v}\pdv{\psi_N(x, t)}{t} + \sigma_t\psi_N(x) + \left(\frac{N}{2N+1}\right)\pdv{\psi_{N-1}(x, t)}{x} + \alpha_N\left(\frac{N + 1}{2N + 1}\right) \pdv{ \psi_{N-1}(x, t)}{x} = 0
\end{equation}
Recalling the definition of $\alpha_N$ in Eqs.~\ref{alphapomraning0} and~\ref{alphapomraning}, we replace $\psi_a$ to be the time-dependent asymptotic solution of the mono-energetic time-dependant (or $s$-dependent) Boltzmann equation introduced in Eq.~\ref{TD_asymptotc_eq}. Thus, our new $\alpha_N$ expression is $s$-dependant and is given by:
\begin{equation}\label{alpha_s}
    \alpha_N\to\halpha{}_{,N} = \frac{\int_{-1}^{1} \hpsi{}_{,a}(x,\mu) P_{N+1}(\mu) d\mu}{\int_{-1}^{1} \hpsi{}_{,a}(x,\mu) P_{N-1}(\mu) d\mu}= \frac{\int_{-1}^{1} \left(\frac{1}{1 + \mu \hkappa}\right) P_{N+1}(\mu) d\mu}{\int_{-1}^{1} \left(\frac{1}{1 + \mu \hkappa}\right) P_{N-1}(\mu) d\mu}
\end{equation}
Substituting the new $\halpha{}_{,N}$ expression (Eq.~\ref{alpha_s}) in to the last approximated equation (Eq.~\ref{atd_nequation}) yields:
\begin{equation}\label{atd_nequation2}
        \frac{1}{v}\pdv{\psi_N(x, t)}{t} + \sigma_t\psi_N(x, t) + \left(\frac{N + (N+1)\halpha{}_{,N}}{2N+1}\right) \pdv{ \psi_{N-1}(x, t)}{x} = 0
\end{equation}
Secondly, we apply the Laplace transform to Eq. \ref{atd_nequation2} using the $s$-dependent cross section (introduced in Eq.~\ref{Hnotation}):
\begin{equation}\label{atd_nequation_laplace1}
    \hsigma_t\hat{\psi}_{s,N}(x) + \left(\frac{N + (N+1)\halpha{}_{,N}}{2N+1}\right)\pdv{\hat{\psi}_{s,N-1}(x)}{x} = 0
\end{equation}
As one can notice, this equation takes the form of a Fick's law with a new modified $s$-dependent ``diffusion" coefficient defined in the following matter:
\begin{equation}\label{asymptotic_pn_ficklaw}
        \hat{\psi}_{s,N}(x) = -\hDN(x) \pdv{\hat{\psi}_{s,N-1}(x)}{x} \quad\longrightarrow\quad \hDN(x) = \frac{N+(N+1)\halpha{}_{,N}}{(2N+1)\hat{\sigma}_t^s}
\end{equation}

As a generalization of Heizler's asymptotic $P_1$ approximation (Eq.~\ref{shay_secondAB}), we seek to approximate the last $N$ equation, using two new coefficients $\A_N(c)$ and $\B_N(c)$, with the following form: 
\begin{equation}\label{atd_nequation_laplace2}
    \frac{\A_N(c)}{v} \pdv{\psi_N(x, t)}{t} + \B_N(c) \sigma_t \psi_N(x,t) + \pdv{\psi_{N-1}(x,t)}{x} = 0
\end{equation}
Applying the Laplace transform to the above approximated closure equation yields:
\begin{equation}\label{atd_nequation_laplace3}
    \hpsi{}_{,N}(x) = - \frac{v}{s\A_N(c) + v\sigma_t\B_N(c)} \cdot \pdv{\hpsi{}_{,N-1}(x)}{x}
\end{equation}
Substituting the approximate Eq.~\ref{atd_nequation_laplace3} into the exact closure Eq.~\ref{atd_nequation_laplace1} (both of them in the Laplace domain) yields:
\begin{equation}
\frac{(2N+1)(v\sigma_t+s)}{N+(N+1)\halpha{}_{,N}\left(\hkappa(\hc)\right)}=\frac{(2N+1)(v\sigma_t+s)}{N+(N+1)\halpha{}_{,N}\left(\hat{D}_0(\hc)\right)}\approx v\sigma_t\B_N(c) + \A_N(c)s + {\cal O}(s^2)
\label{AnBn}
\end{equation}
i.e., we can solve the modified asymptotic coefficients, $\A_N(c)$ and $\B_N(c)$ inside Eq.~\ref{AnBn} using the integral notation for $\halpha{}_{,N}$, Eq.~\ref{alpha_s} under small-$s$ restrictions (asymptotic on time, using Taylor series in $s$). $\halpha{}_{,N}$ is an explicit function of $\hkappa(\hc)$ for each $\hc$ which has several fits (alternatively we can use the more convenient fits for $\hat{D}_0(\hc)$, which connects directly to $\hkappa(\hc)$ through Eq.~\ref{D0}).
Therefore, choosing the closure of Eq.~\ref{atd_nequation_laplace2}, underlies the new approximation for general $P_N$ closure, or simplified $P_N$ ($SP_N$) in multi-dimensions~\cite{spn}, based on the asymptotic distribution of both space (as Pomraning's time-independent case) and time (using small $s$).

Next, we proceed in the explicit calculation of the variables $\A_N$ and $\B_N$ for $N=1,2,3$, afterwards we present the derivation for general $N$.

\subsection{The First Three Time-dependent Asymptotic $P_N$}\label{three_tda_pn}
In this section we derive the first three {\em asymptotic} time-dependent $P_N$ equations. As explained, deriving the $N$ approximated equation is done by: I: Calculating $\halpha$. II: Obtaining the diffusion-like coefficient $\hD{}_{,n}$. III: Expanding the expression of $\frac{v}{\hD{}_{,n}}$ in a Taylor series. IV: Calculating the variables $\A_N$ and $\B_N$.
\subsubsection{Time-dependent Asymptotic $P_1$}

Firstly, deriving the asymptotic $P_1$ equation, as a kind of sanity check since the expression should be exactly the same as Heizler's $P_1$ approximation~\cite{heizler2010asymptotic}.
Calculating $\halpha{}_{,1}$ for the closure equation yields:
\begin{equation}\label{alpha_s1}
    \halpha{}_{,1}=\frac{\int_{-1}^1 P_2(\mu)\frac{1}{1 + \mu\hkappa} d\mu}{\int_{-1}^1 P_0(\mu)\frac{1}{1 + \mu\hkappa} d\mu} = \frac{3(1-\hc)}{2\hkappa^2} - \frac{1}{2}
\end{equation}
Next, calculating the new modified diffusion-like coefficient yields:
\begin{equation}
    \hD{}_{,1} = \frac{1+2\halpha{}_{,1}}{3\hat{\sigma}_t^s} = \frac{(1-\hc)}{\hkappa^2\hsigma_t}
\end{equation}
Thus, the last equations is:
\begin{equation}\label{asymptotic_p1_ficklaw}
        \hat{\psi}_{s,1}(x) = -\frac{(1-\hc)}{\hkappa^2 \hsigma_t} \pdv{\hat{\psi}_{s,0}(x)}{x}
\end{equation}
Which is indeed, {\em exactly} as in Eq.~\ref{heizler_p1}~\cite{heizler2010asymptotic}. Thus, following the same procedure as described and performed for the asymptotic $P_1$ approximation, we can solve for $\A_1$ and $\B_1$.

For example, for the case of $(1-c) \ll 1$, using the appropriate $D_0$ from the appendix (Eq.~\ref{D0_big_c}) one can calculate $\A_1$ and $\B_1$:
\begin{equation}\label{a1b1_atd}
       \frac{v}{\hD{}_{,1}} = \frac{v\sigma_t + s}{D_0(\hc)} \approx \frac{v\sigma_t + s}{\frac{1}{3}(1+\frac{4}{5}(1-\frac{cv\sigma_t}{v\sigma_t + s}))} \approx \frac{15v\sigma_t}{9-4c} + \frac{15 (9-8c)s}{(9-4c)^2}
\end{equation}
Thus we achieve the following expressions for the case of a highly scattering case: $\A_1=\frac{15(9-8c)}{(9-4c)^2}$ and $\B_1 = \frac{15}{9-4c}$. For $c=1$ (purely scattering case) $\A_1=3/5$ and $\B_1=3$ exactly, reproducing the asymptotic $P_1$ approximation of Heizler~\cite{heizler2010asymptotic}.
For general $c$, one can use the different approximate fits of $D_0(c)$ (see the appendix), solving for $\A_1$ and $\B_1$.
In Fig.~\ref{fig:AB} the variables $\A_1(c)$ and $\B_1(c)$ are shown in green, which reproduces Heizler's asymptotic $P_1$ for any given $c$~\cite{heizler2010asymptotic,heizler2012sp,cohen2018discontinuous}.
\begin{figure}[htbp!]
\includegraphics*[width=7.5cm]{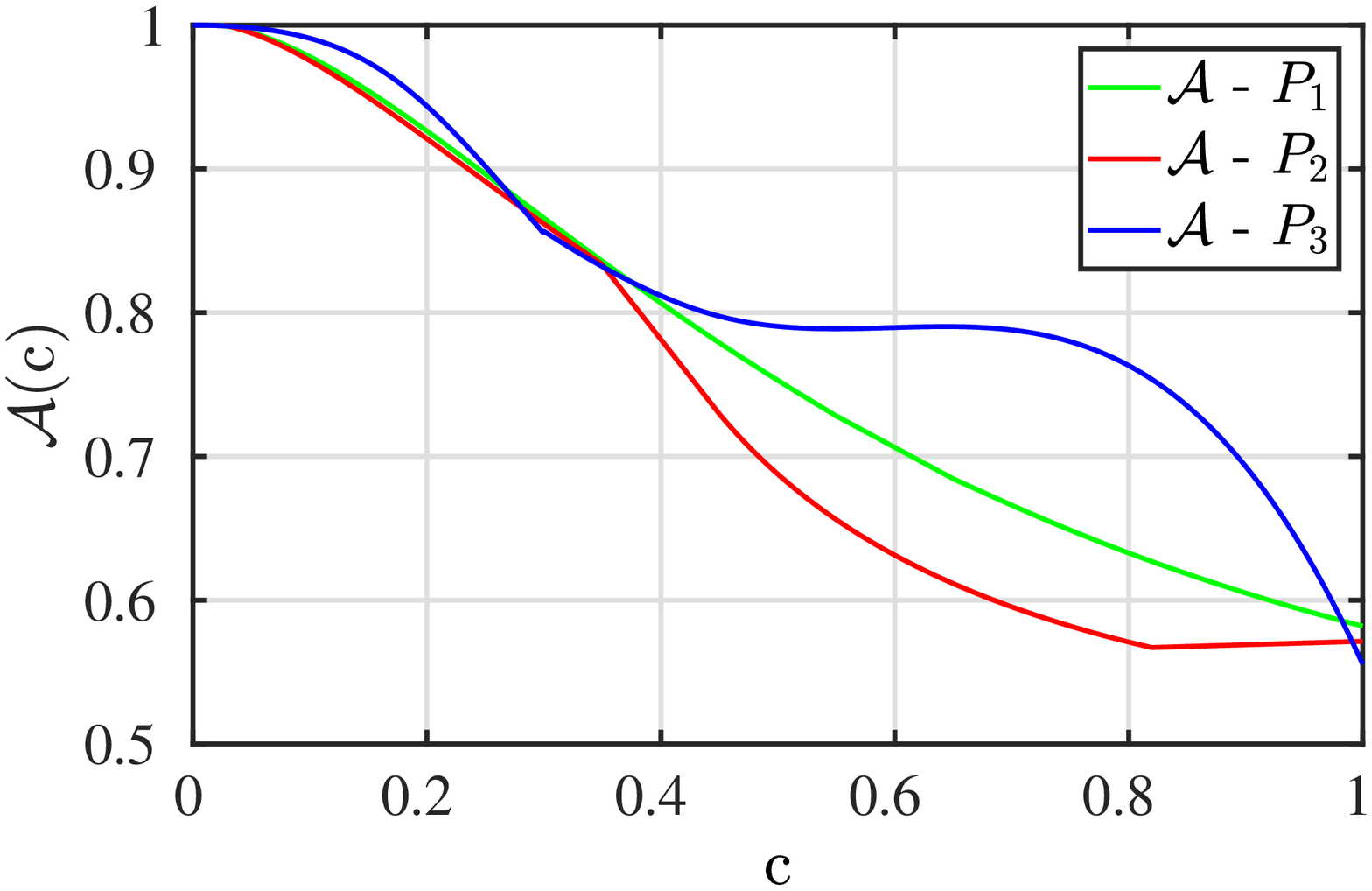}
\includegraphics*[width=7.5cm]{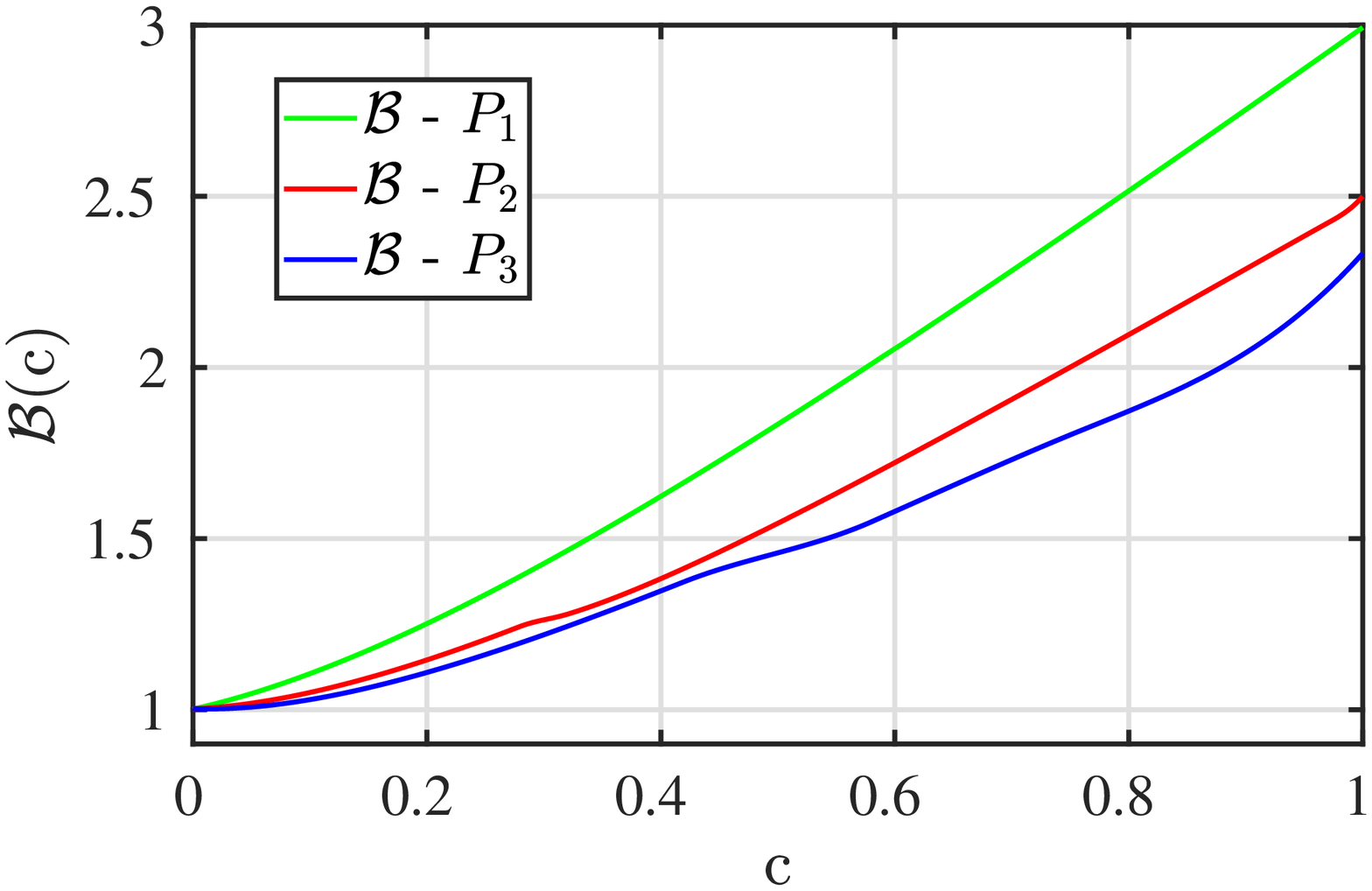}
\caption{The media-dependent coefficients $\A_N(c)$ and $\B_N(c)$ as function of the medium's albedo $c$ for the first 3 orders of the time-dependent asymptotic $P_N$ approximation ($N=1,2,3$).}
\label{fig:AB}
\end{figure}

\subsubsection{Asymptotic $P_2$}
Deriving asymptotic $P_2$ is done by following the same procedure explained above, for $N=2$. First $\halpha{}_2$ is calculated:
\begin{equation}\label{alpha_s2}
     \halpha{}_{,2} = \frac{9\hkappa^2 - 15 + \hc(15-4\hkappa^2)}{-6\hkappa^2(1-\hc)}
\end{equation}
And the diffusion coefficient:
\begin{equation}
    \hD{}_{,2} = \frac{3(1 -\hc)-\hkappa^2}{2\hkappa^2\hsigma_t(1-\hc)}=\frac{(1-3\hat{D}_0(\hc))}{2\hsigma_t(1 - \hc)}
\label{ap2_D}
\end{equation}
As an additional sanity check, we use the expression for $D_0(c)$ of the {\em classic $P_2$ approximation}, $D_0(c) = \frac{9-4c}{15}$ into Eq.~\ref{ap2_D} (It is well-known that the $P_2$ approximation can be introduced in classic Fick's law with this modified diffusion coefficient, in the absence of an external source~\cite{ShinMorel1993,PomraningRulkoSu1994,Rulko1995,heizler2012sp}). Using this classic $P_2$ diffusion coefficient, we indeed reproduce the classic $P_2$ equations, i.e. $\A_2 = \B_2 = 5/2$, for all given $c$.

Back to the asymptotic derivation, substituting into Eq.~\ref{ap2_D} the asymptotic expression of $D_0(c)$ (see the appendix), for example for $(1-c) \ll 1$ (highly scattering medium), and expanding the expression in a Taylor series, one obtains: 
\begin{equation}
    \frac{v}{\hD{}_{,2}} = \frac{875 (s+v\sigma_t)}{818 + \frac{198c^2}{(1+\frac{s}{v\sigma_t})^2} - \frac{666c}{1+\frac{s}{v\sigma}}}\approx \frac{875v\sigma_t}{818 -666c +198c^2} + \frac{875(409 -666c +297c^2)s}{2(409-33c+99c^2)^2} + {\cal O}(s^2)
\label{ap2exp}
\end{equation}
From Eq.~\ref{ap2exp}, $\A_2$ and $\B_2$ are found:
\begin{subequations}
    \begin{equation}
    \A_2 = \frac{875(409 -666c +297c^2)}{2(409-33c+99c^2)^2}
\end{equation}
\begin{equation}
    \B_2 = \frac{875}{818 -666c +198c^2}
\end{equation} 
\end{subequations}
For pure scattering medium $c=1$, we obtain $\A_2=4/7$ and $\B_2=5/2$. For general $c$, using the different approximate fits of $D_0(c)$ (see the appendix), the coefficients $\A_2$ and $B_2$ are shown in Fig.~\ref{fig:AB} in red.

\subsubsection{Asymptotic $P_3$}
Deriving the asymptotic time-dependent for $N=3$, $P_3$ follows again, the same procedure:
\begin{equation}\label{alpha_s3}
    \halpha{}_{,3}= - \frac{105(1-c) -90\hkappa^2 + 9\hkappa^4 + 55\hkappa^2\hc}{12\hkappa^2(3(1-\hc) - \hkappa^2)}
\end{equation}
and the new modified Fick's law coefficient:
\begin{equation}
    \hD{}_{,3} = \frac{9\hkappa^2 - 15 + \hc(15-4\hkappa^2)}{3\hkappa^2\hsigma_t(\hkappa^2 + 3\hc - 3)} = \frac{\hat{D}_0(\hc) (4\hc + 15\hat{D}_0(\hc) - 9)}{3 (3\hat{D}_0(\hc) - 1) (1-\hc) \hsigma_t}
\end{equation}

Again, for highly scattering medium ($(1-c) \ll 1$) using Eq.~\ref{D0_big_c} yields:
\begin{equation}
    \frac{v}{\hD{}_{,3}} = \frac{175 (409 + \frac{99c^2}{(1+\frac{s}{v\sigma})^2})(s+v\sigma)}{(-26+\frac{11c}{(1+\frac{s}{v\sigma})}) (-2511 + \frac{396c^3}{(1+\frac{s}{v\sigma})^2} - \frac{1728c^2}{(1+\frac{s}{v\sigma})^2} +\frac{2968c}{(1+\frac{s}{v\sigma})})} 
\label{exact_ap3}
\end{equation}
Expanding Eq.~\ref{exact_ap3} in a Taylor series yields:
\begin{align}
& \frac{v}{\hD{}_{,3}} \approx
\frac{175(409 -333c +99c^2)v\sigma}{(11c-26) (-2511+2968c -1728c^2 +396c^3)} + \\
&\quad \frac{175(26701974 -85717402c + 123617175c^2 - 99606960c^3 + 45862740c^4 - 11604384c^5 + 1293732c^6}{(11c-26)^2 (-2511+2968c -1728c^2 +396c^3)^2}s     \nonumber
\end{align}

Obtaining once more the $\A_3$ and $\B_3$ coefficients for purely scattering medium ($c=1$) yields $\A_3=5/9$ and $\B_3 =7/3$. For general $c$, using the different approximate fits of $D_0(c)$ (see the appendix), the coefficients $\A_3$ and $B_3$ are shown in Fig.~\ref{fig:AB} in blue.

Fig.~\ref{fig:AB} suggests some important behaviors: First, $\B_N$ always equals to the inverse of Pomraning's time-independent asymptotic $P_N$ ``diffusion-like" coefficient, $\B_N=1/D_N(c)$, as expected. This forces the solution of the time-dependent asymptotic $P_N$ to tend the solution of the time-independent asymptotic $P_N$ of Pomraning, when $t\to\infty$. Second, for pure scattering medium ($c=1$), $\B_N$ equals to the classic $P_N$ approximation coefficient for general $N$ but $\A_N$ is different (as in Heizler's asymptotic $P_1$ has shown~\cite{heizler2010asymptotic,heizler2012sp}). Third, for pure absorbing case ($c=0$), $\A_N(0)=\B_N(0)=1$ for any general $N$, yielding the exact particle velocity.

\subsection{General $N$ derivation}
In this section we generalize solution for $\halpha{}_{,N}$.
For simplicity, we will express the Legendre polynomial $P_N$ in a simple sum series of polynomials:
\begin{equation}\label{simple_pn}
    P_N(\mu) = \sum_{n=0}^{N} a_n \mu^n
\end{equation}
Where the coefficients of the Legendre polynomial $a_n$ are given by:
\begin{equation}
a_n= 2^N \binom{N}{n} \binom{\frac{N+n-1}{2}}{N}
\end{equation}
Substituting Eq.~\ref{simple_pn} into Eq.~\ref{alpha_s} yields a new form of the integrals in Eq.~\ref{alpha_s} which will be expressed for general $N$. The integral of the numerator takes the following form:
\begin{align}\label{solution_integral}
    & \int_{-1}^1 \frac{P_{N+1}(\mu)}{1+\mu\vk}d\mu = \sum_{n=0}^{N+1} a_n \int_{-1}^1 \frac{\mu^nd\mu}{1+\mu\hkappa} = \\
    & \sum_{n=0}^{N+1} a_n \frac{\hF(1,n + 1;n + 2;-\hkappa) - (-1^{n+1}) \hF(1,n+1;n+2;\hkappa)}{n + 1} \nonumber
\end{align}
where $\hF (1, n + 1; n +2; -\hkappa)$ is the hypergeometric function. The values of this specific hypergeometric function is:
\begin{equation}
    \hF(1,n+1;n+2;\hkappa) = -\frac{n+1}{\hkappa^{n+1}}\left(\ell n(1-\vk) + \sum_{k=1}^{n} \frac{\vk^k}{k} \right)
\label{hyper}
\end{equation}
Substituting Eq.~\ref{hyper} into Eq.~\ref{solution_integral} and simplifying using Eq.~\ref{hkappa}, the general solution for the integral takes this form:
\begin{equation}\label{pn_integral_solve}
        \int_{-1}^1 \frac{P_{N+1}(\mu)}{1+\mu\hkappa}d\mu = -\sum_{n=0}^{N+1} 
            \frac{a_n}{\hkappa^{n+1}} \left((-1)^{n+1}\frac{2\hkappa}{\hc} + \sum_{k=1}^{n} \frac{\hkappa^k ((-1)^{n} + (-1)^{k - (n+1)})}{k}\right)
\end{equation}
It is worth noting that for even $k$ (and due to the parity of the Legendre polynomials), the expression $\left(\hkappa^k ((-1)^{n} + (-1)^{k - (n+1)}\right)/k$ cancels out. Thus, only odd $k$ will contribute to the sum. The solution for the denominator in Eq.~\ref{alpha_s} (with an index $N-1$) can be obtained in the same manner as the solution for the index $N+1$.

Deriving $\halpha{}_{,N}$ for $N=1,2,3$ yields the exact result presented in section~\ref{three_tda_pn}. 

\subsubsection*{Alternative derivation}
Alternative derivation of the integrals in Eq.~\ref{alpha_s} for evaluating $\halpha{}_{,N}$ may be done in a different manner of using Eqs.~\ref{solution_integral} and~\ref{pn_integral_solve} and the hypergeometric functions. In this alternate derivation, Pomraning~\cite{pomraning2005equations} expresses the definition of the asymptotic $\halpha{}_{,N}$ as:
\begin{equation}
    \halpha{}_{,N} = \frac{Q_{N+1}\left(\frac{1}{\hkappa}\right)}{Q_{N-1}\left(\frac{1}{\hkappa}\right)}
\label{alter}
\end{equation}

where $Q_n$ is the Legendre function of the second kind and is defined by the following identity:
\begin{equation}
Q_n(z) = \frac{1}{2}\int_{-1}^{1}d\mu \frac{P_n(\mu)}{z-\mu} \xrightarrow{z=\frac{1}{\hkappa}} Q_{n}\left(\frac{1}{\hkappa}\right) \equiv \frac{1}{2}\int_{-1}^{1}d\mu \frac{\hkappa P_{n}(\mu)}{1 - \mu \hkappa}
\end{equation}

Substituting $\hkappa$ inside Eq.~\ref{alter}, one can expand $\halpha{}_{,N}$ using the recursion formula of the Legendre functions of the second kind~\cite{legendre}:
\begin{equation}
Q_n(x)=\begin{cases}
         \frac{1}{2}\ell n\left(\frac{x+1}{x-1}\right) & 
        n=0 \\ 
        P_1(x)Q_0(x)-1 & 
        n=1 \\
        \frac{2n-1}{n}xQ_{n-1}(x)-\frac{n-1}{n}xQ_{n-2}(x) &
        n\geqslant 2
        \end{cases}
\end{equation}
and the definition of $\hkappa$, using Eq.~\ref{hkappa}.

For example, substituting in Eq.~\ref{alter} $N=1$, $\halpha{}_1$ takes this form:
\begin{equation}\label{alpha1_2kind}
    \halpha{}_{,1} = \frac{Q_2(1/\hkappa)}{Q_0(1/\hkappa)} = \frac{1}{2}\frac{3\ell n\left(\frac{1-\hkappa}{1+\hkappa}\right)-
    \hkappa^2\ell n\left(\frac{1+\hkappa}{1-\hkappa}\right) - 6\hkappa}{\hkappa^2\ell n\left(\frac{1+\hkappa}{1-\hkappa}\right)}
\end{equation}
Recalling the closed transcendental Eq.~\ref{hkappa} for $\hkappa$, and substituting it in the Eq.~\ref{alpha1_2kind} results:
\begin{equation}
   \halpha{}_{,1} = \frac{1}{2} \frac{3\frac{2\hkappa}{\hc}-\hkappa^2 \frac{2\hkappa}{\hc} - 6\hkappa}{\hkappa^2 \frac{2\hkappa}{\hc}} = \frac{3(1-\hc)}{2\hkappa^2} - \frac{1}{2},
\end{equation}
which is identical to the result shown in Eq.~\ref{alpha_s1}.

Following the same procedure for $N=2$:
\begin{equation}\label{alpha2_2kind}
 \halpha{}_{,2} = \frac{Q_3(1/\hkappa)}{Q_1(1/\hkappa)} = \frac{1}{6}\frac{15\ell n\left(\frac{1+\hkappa}{1-\hkappa}\right) -9\hkappa^2\ell n\left(\frac{1+\hkappa}{1-\hkappa}\right) + 8 \hkappa^3 -30\hkappa}{\hkappa^2\ell n\left(\frac{1+\hkappa }{1-\hkappa}\right) - 2\hkappa^3}
\end{equation}
Recalling the closed transcendental Eq.~\ref{hkappa}, and substituting it in the Eq.~\ref{alpha2_2kind} results:
\begin{equation}
    \halpha{}_{,2} = \frac{\frac{30\hkappa}{\hc} -\frac{18\hkappa^3}{\hc} +8\hkappa^3 -30\hkappa}{6\hkappa^3\left(\frac{2}{\hc} - 2\right)} = \frac{15 - 9 \hkappa^2 +4\hkappa^2\hc -15 \hc}{6\hkappa^2(1-\hc)},
\end{equation}
which is the same as Eq.~\ref{alpha_s3}. The rest of the derivation is the same as shown in Sec.~\ref{heart}.

\section{Green-Function Benchmark}\label{test_benchmark}

In order to test our time-dependent asymptotic $P_N$ approximation, we used the benchmark known as the AZURV1 benchmark~\cite{ganapol1999,olson2004numerical}. The AZURV1 benchmark is a full green function - both in time and space ($Q(x,t) = Q_0 \delta(x)\delta(t)$) for the one-dimensional infinite homogeneous media Boltzmann equation (Eq.~\ref{transport_Eq}).
The results presented, show the propagation of the normalized scalar flux distribution as a function of dimensionless space and time. The numerical results of the new approximation, were compared to approximations presented in section~\ref{section_pn}.
In addition to the semi-analytic results of the AZURV1 benchmark, we include numerical simulations of time-dependent $S_N$, with $N=64$, as a good approximation of the exact results (when $N\to\infty$, the $S_N$ method, which is equivalent to a $P_{N-1}$ calculation, tends to the exact Boltzmann equation), yielding a continuous trend of the exact benchmark.

The $S_N$ code was written using fully implicit scheme on time and diamond difference scheme with negative-flux-fixup in space~\cite{BellGlasstone} in {\em{Fortran}} using a constant $ \sigma_t\Delta x\approx 6\cdot 10^{-5}$ while $v\sigma_t \Delta t$ is defined dynamically such that the particles' flux ($\phi$) will not change in each cell (between time steps) more than 0.5\%.  
For all of the $P_N$ approximations (both classic and asymptotic), numerical simulations were executed in a fully implicit scheme on time with a finite difference scheme on space in {\em{Matlab}} (we have used the \textit{sparse} option for accelerating the band-matrix inversion). We used a constant spatial and temporal resolution of $ \sigma_t \Delta x = v\sigma_t \Delta t = 1.25\cdot 10^{-3}$.

\subsection{Pure Scattering Case}
The first scenario that we use to test our approximation is the purely scattering case ($c=1$). The pure scattering case tends to the diffusion limit faster than all other cases at a given time, as it has no absorbing terms and will behave like diffusion. This case is important also because the exact transport solution for the Green function in the slab geometry yields a scaling relation of this form~\cite{Kuscer1965,CaseZweifel1967}:
\begin{equation}
\phi^{(c)}(x,t)=ce^{-(1-c)v\sigma_tt}\phi^{(1)}(cx,ct)
\label{scale}
\end{equation}
This means, knowing the solution for $c=1$ and $\phi^{(1)}(x,t)$, yields the exact solution for general $c$ and $\phi^{(c)}(x,t)$. Thus, in case of a multiplying medium ($c>1$) and the case of absorbing medium ($c<1$) the solution still holds.
\begin{figure}[htbp!]
\includegraphics[width=18cm,trim={0 0 5cm 0},clip]{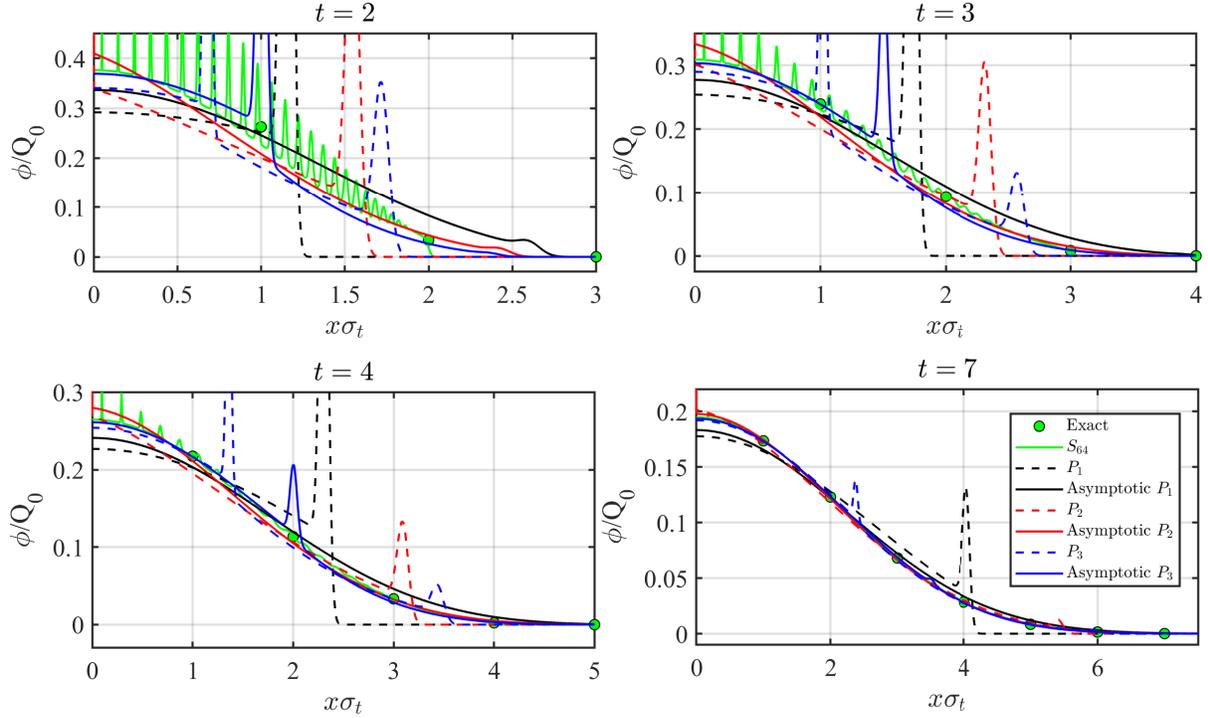}
\caption{The scalar flux of the different approximations and the exact benchmark as a function of dimensionless space for the fully scattering case ($c=1$). The semi-analytic solution is presented in green circles, while the $S_{64}$ results are presented in the solid green curves. The classic $P_1$ is in the dashed black curves, the time-dependent asymptotic $P_1$ is in the solid black curves. The $P_2$ and the $P_3$ are in the red and blue curves, respectively, while dashed is for classic and solid for asymptotic.}
\label{fig:c1_t}
\end{figure}
\begin{figure}[htbp!]
\includegraphics*[width=7.5cm]{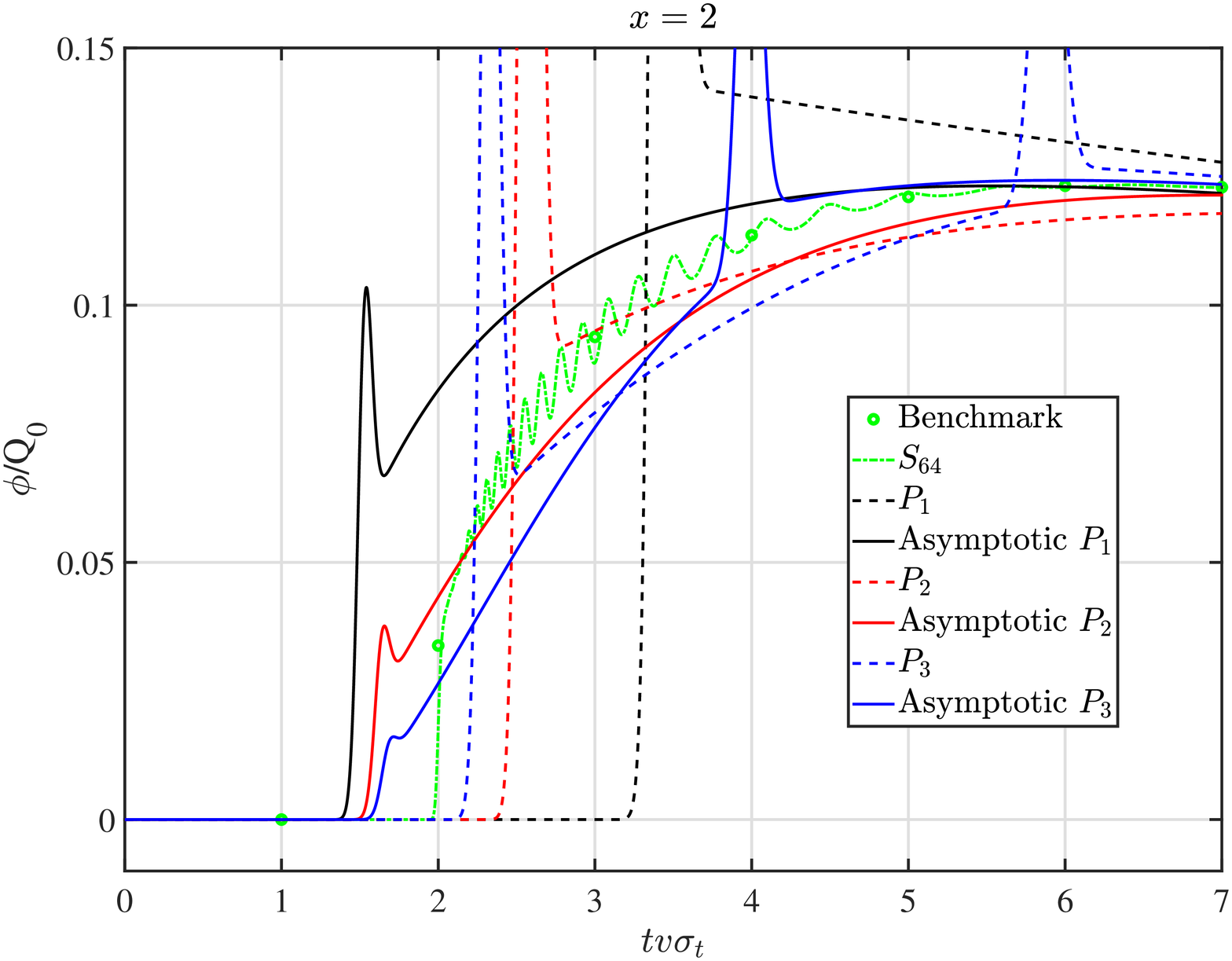}
\includegraphics*[width=7.5cm]{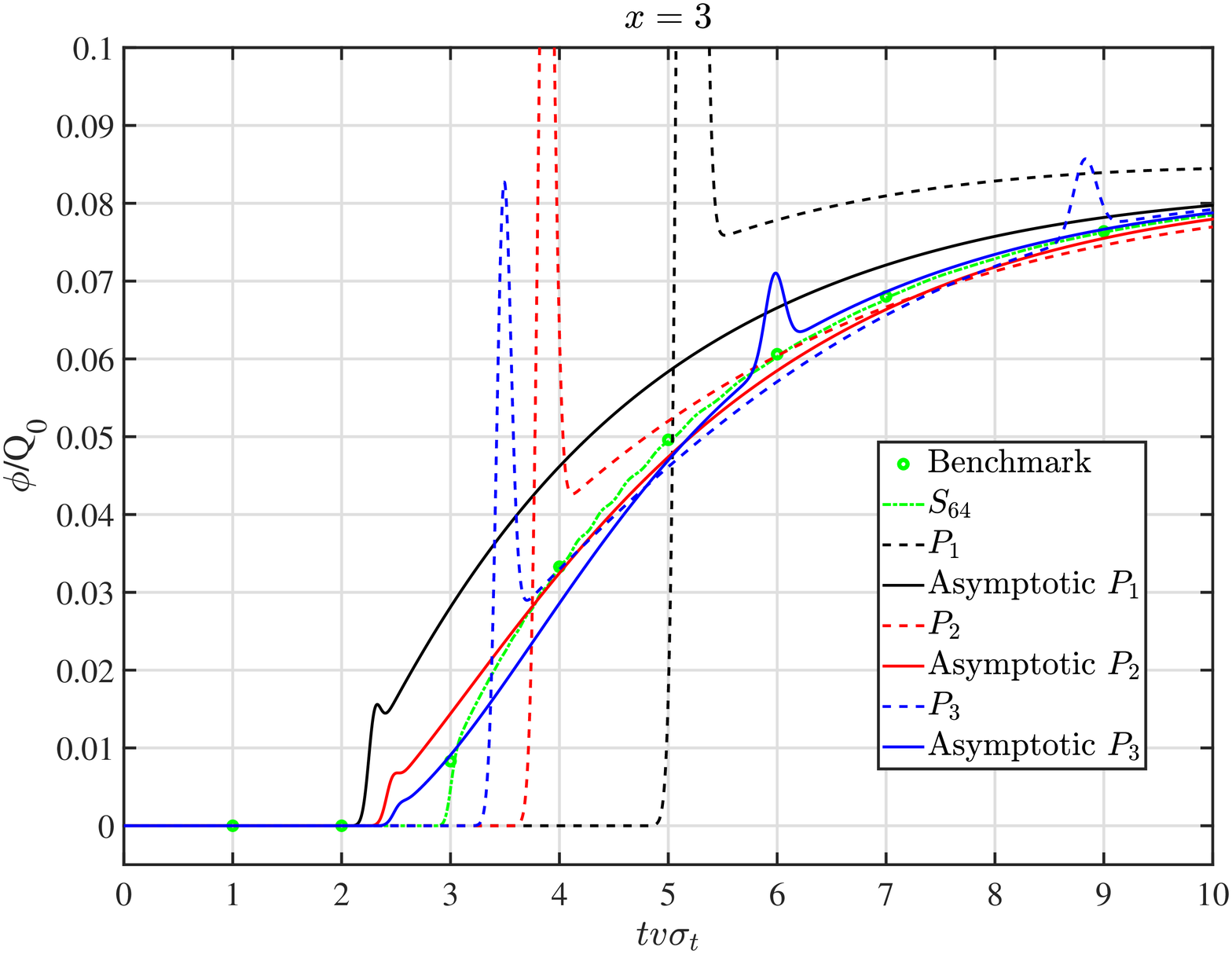}
\includegraphics*[width=7.5cm]{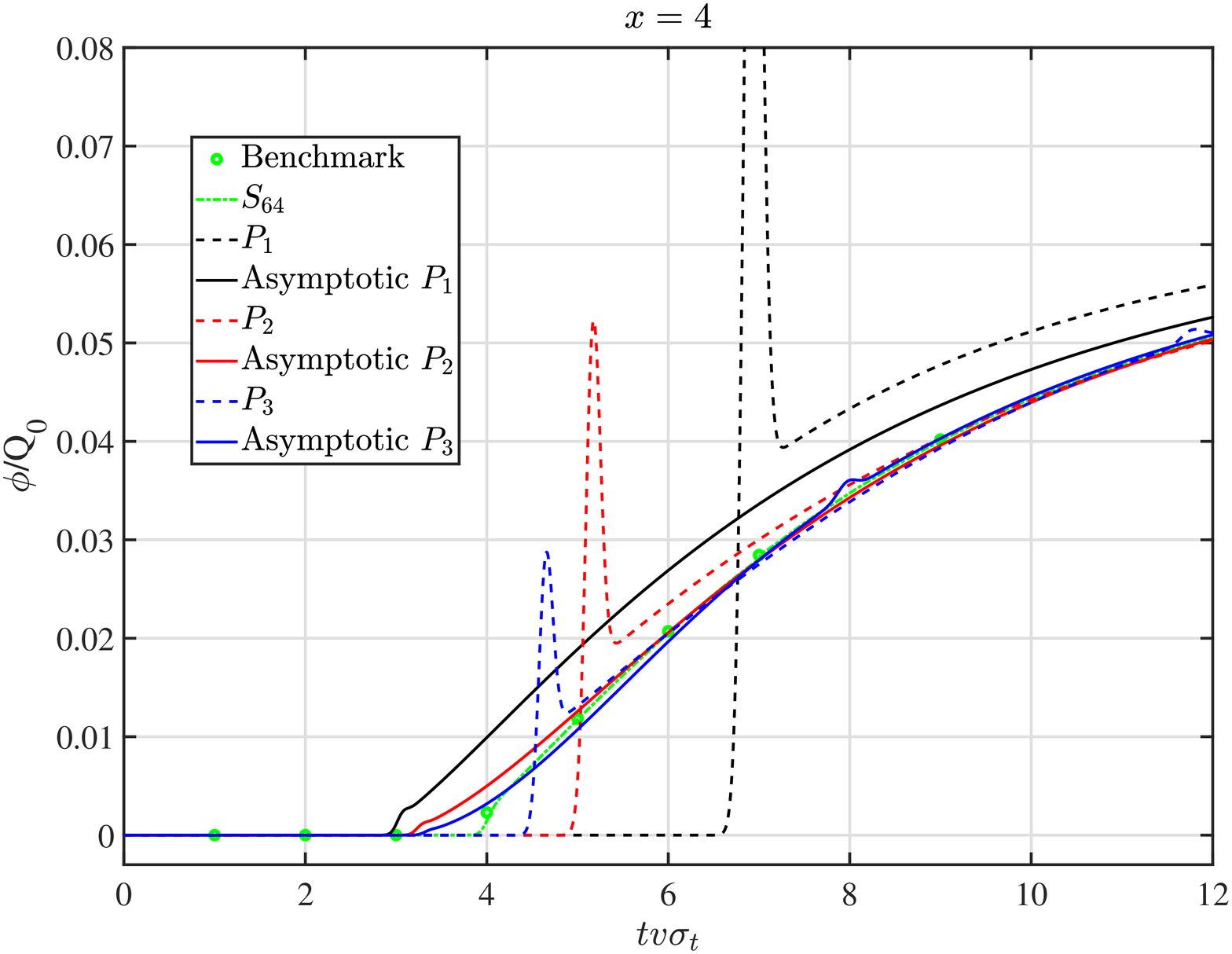} 
\caption{The scalar flux of the different approximations as a function of the dimensionless time for the fully scattering case ($c=1$). The semi-analytic solution is presented in green circles, while the $S_{64}$ results are presented in the solid green curves. The classic $P_1$ is in the dashed black curves, the time-dependent asymptotic $P_1$ is in the solid black curves. The $P_2$ and the $P_3$ are in the red and blue curves, respectively, while dashed is for classic and solid for asymptotic.}
\label{fig:c1_x}
\end{figure}

In Figs.~\ref{fig:c1_t} and~\ref{fig:c1_x} we compare the different approximations presented in the paper in terms of the normalized scalar flux to the source power $Q_0$. In Fig.~\ref{fig:c1_t} the scalar flux is compared at different times ($t = 2,3,4$ and 7) while in Fig.~\ref{fig:c1_x} the scalar flux is compared in different positions ($x=2, 3$ and $4$). The solution is presented in normalized position (in $\sigma_t$ units) and normalized time (in $v\sigma_t$ units).

As one can see, in both figures the $S_{64}$ reproduces the exact benchmark while the ``spikes" represent the different delta function for each discrete direction of the $S_{64}$ (at later times, the delta functions smooth away due to numerical features and re-distribution due to collisions). First, although the asymptotic $P_1$ yields much better results than classic $P_1$ (as known~\cite{bengston1958}), there is still a non-negligible difference between asymptotic $P_1$ results and the exact behavior (see widely in ~\cite{heizler2010asymptotic}).
Both $P_1$ and $P_2$ (both classic or asymptotic) has one spike (that smooths away at later times due to numerical features and re-distribution due to collisions) that represents the speed velocity of the approximation. The $P_3$ approximations have two different spikes (since $P_3$ is equivalent to $S_4$). The classic particle velocities are always slower than the real particle velocity, while the asymptotic $P_N$ converges to the exact velocity from the other direction (faster than the particle speed).

However, the main result that can be seen from Fig.~\ref{fig:c1_t} and especially from Fig.~\ref{fig:c1_x} (that is focused on the tails of the distributions), that the asymptotic $P_N$ approximation yields better results than classic $P_N$ approximation, for any given $N$ (at least in the low-order $N$'s), and thus, converges faster than classic $P_N$. Even for $N=2$, the asymptotic $P_2$ approximation yields very good approximate solution with the exact solution, even near the tail of the distribution. 

\begin{figure}[htbp!]
\includegraphics*[width=7.5cm]{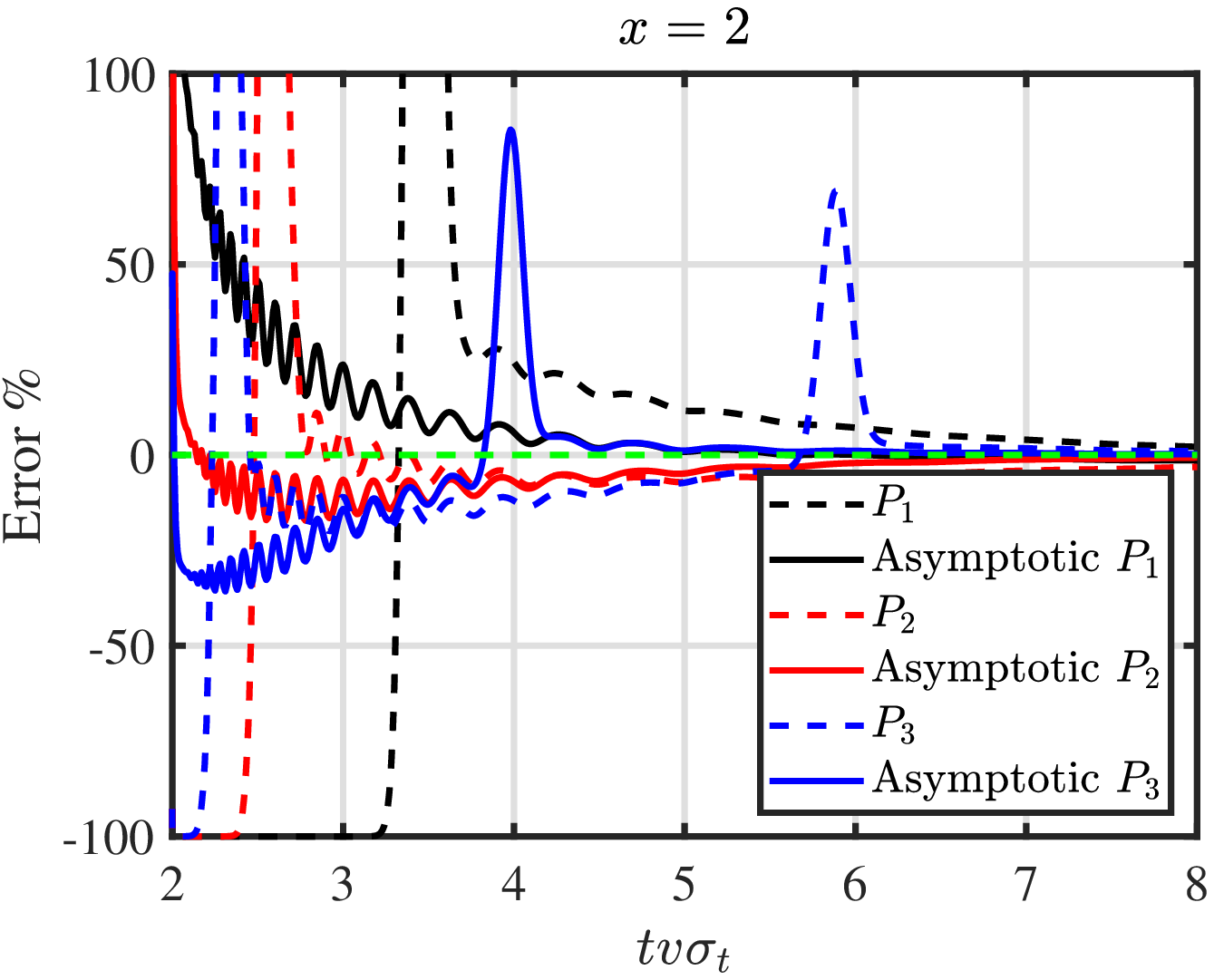}
\includegraphics*[width=7.5cm]{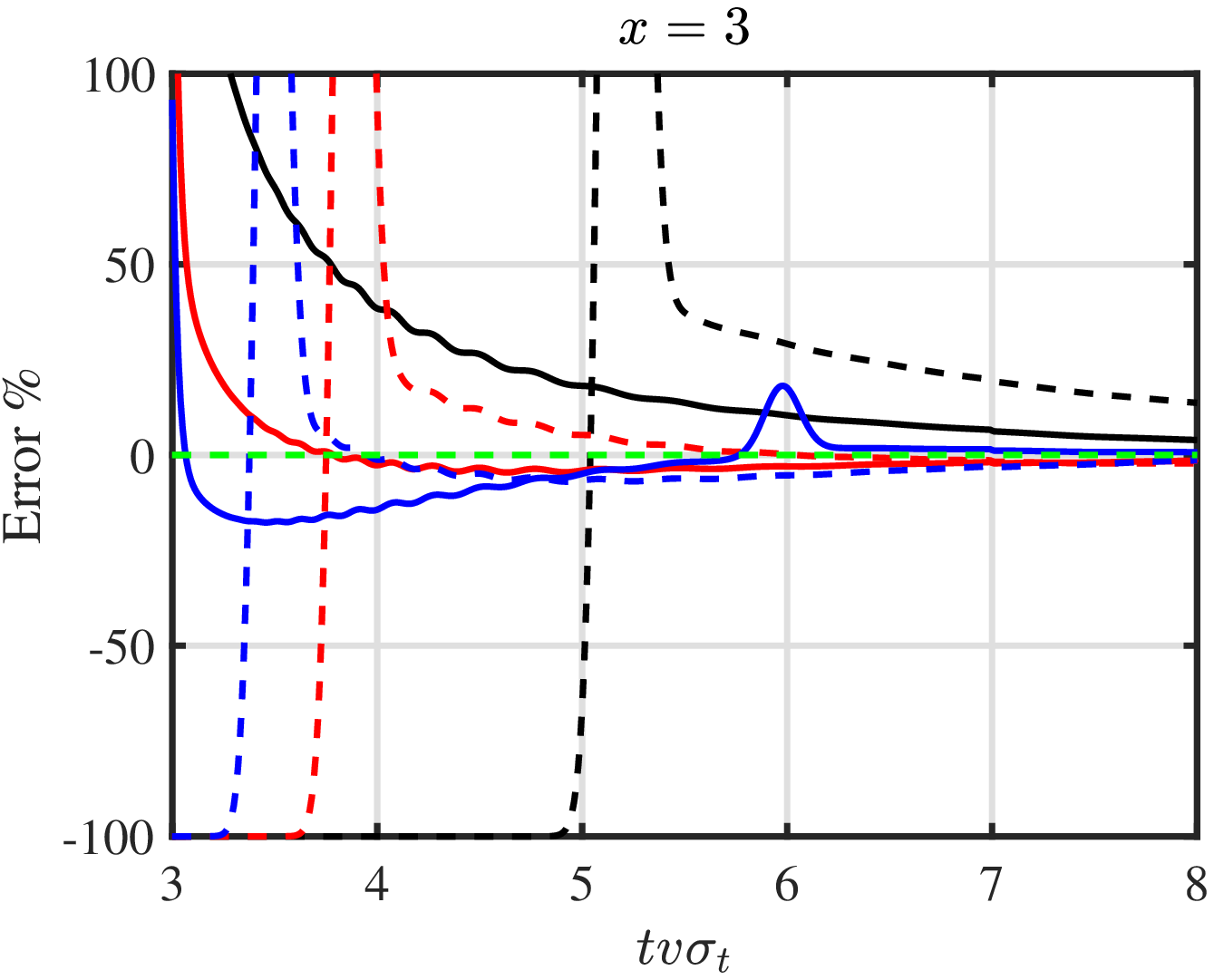}
\includegraphics*[width=7.5cm]{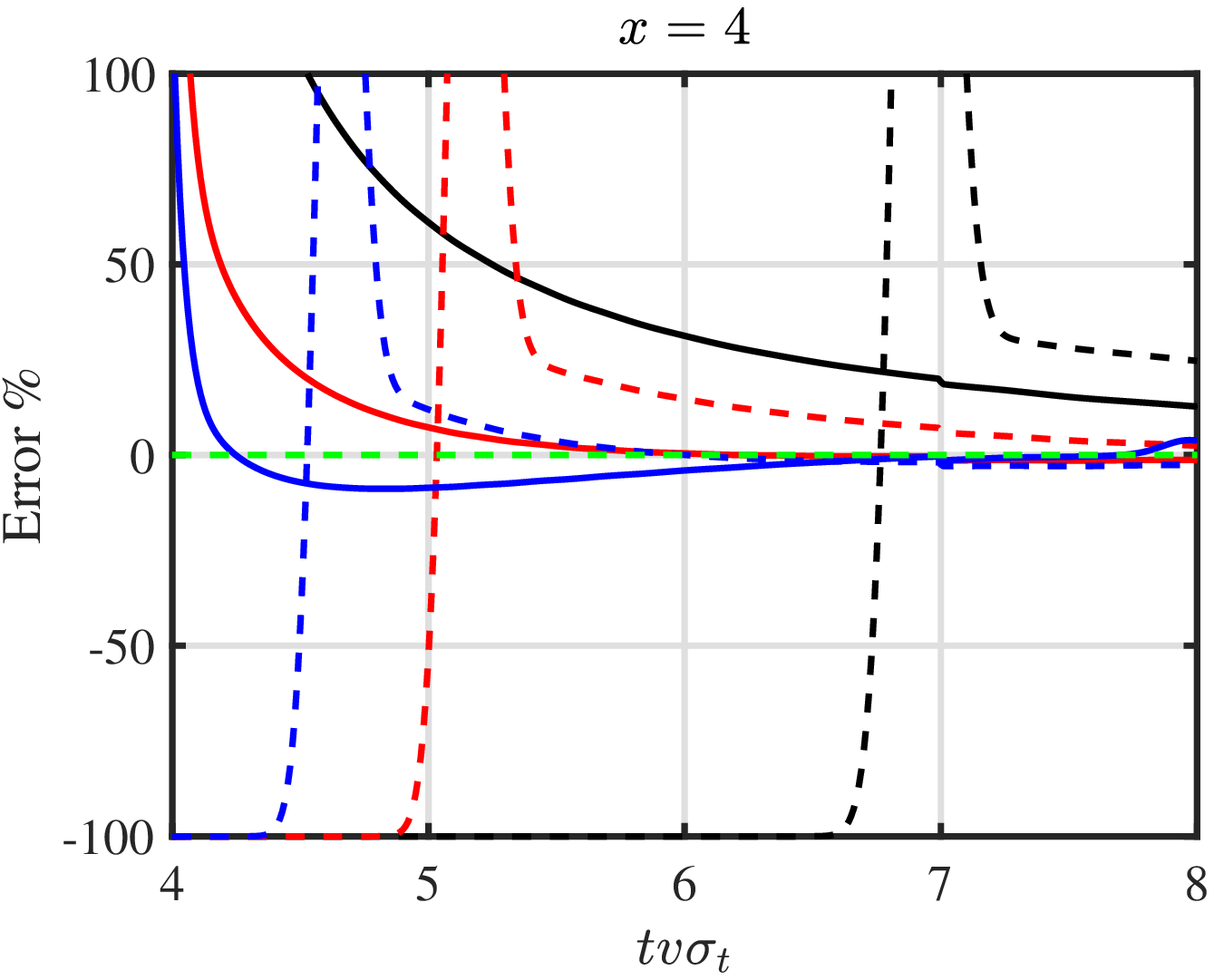}
\caption{The relative error (in \%) between the different $P_N$ approximations to the $S_{64}$ as a function of the dimensionless time for the fully scattering case ($c=1$).}
\label{fig:error_c1_x}
\end{figure}

To emphasize quantitatively the benefit of the time-dependent asymptotic $P_N$ approximation compared to the classic $P_N$ approximation for any given $N$, the relative error between the $P_N$ approximation and the $S_{64}$ (Eq.~\ref{error}) was calculated via the following definition:
\begin{equation}\label{error}
\mathrm{Error}=100 \cdot \frac{\phi_{\mathrm{Approx}}-\phi_{\mathrm{Exact}}}{\phi_{\mathrm{Exact}}},
\end{equation}
and presented in Fig.~\ref{fig:error_c1_x}.

We can notice that for any given $N$, the relative error of the time-dependent asymptotic $P_N$ is smaller than its counter approximation - the classical $P_N$. For $P_1$ (black curves) the benefit is clear. In the case of $N=2$ (red curves), the error of the time-dependent asymptotic $P_N$ is almost flat, until it rises near the front itself (at $x=vt$). Finally for the case of $N=3$, both approximation yield relatively small errors, but the time-dependent asymptotic $P_N$ yields better results, especially near the tails of the distribution, and in the asymptotic regimes ($t\to\infty$).

\subsection{Asymptotic $P_N$ vs. other modified closures}

\begin{figure}[htbp!]
\includegraphics*[width=13cm]{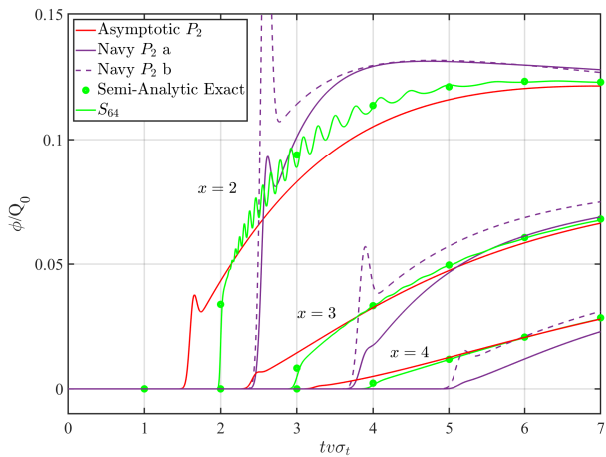}
\caption{A comparison between the asymptotic $P_2$ approximation to Dawson's (navy) modified $P_2$ approximation. The scalar flux of the different approximations as a function of the dimensionless time for the fully scattering case ($c=1$). Dawson's modified $P_2$ approximations are in the violet curves. ``Navy $P_2$ a" represents $\alpha_1 = 1.1.8858$ and $\alpha_2=4.16349$ and ``Navy $P_2$ b" represents $\alpha_1 = 1$ and $\alpha_2=2.4$.}
\label{fig:navy}
\end{figure}

To further test the results of the time-dependent asymptotic $P_2$ we compared it to a time-dependent version of the two options that Dawson proposed in his modified $P_2$ approximation~\cite{dawson} (see Sec.~\ref{dawson_section}). The comparison is presented for $x=2,3$ and 4 as a function of time in Fig.~\ref{fig:navy}, when the modified $P_2$ approximations are the violet curves. 

One can see that as Dawson marks~\cite{dawson}, option ``Navy $P_2$ a`` yields better results than ``Navy $P_2$ b``. However, the time-dependent asymptotic $P_2$ yields better results than both of the modified $P_2$ Dawson's proposals, both in the bulk and especially for the tails.

\begin{figure}[htbp!]
\includegraphics*[width=7.5cm]{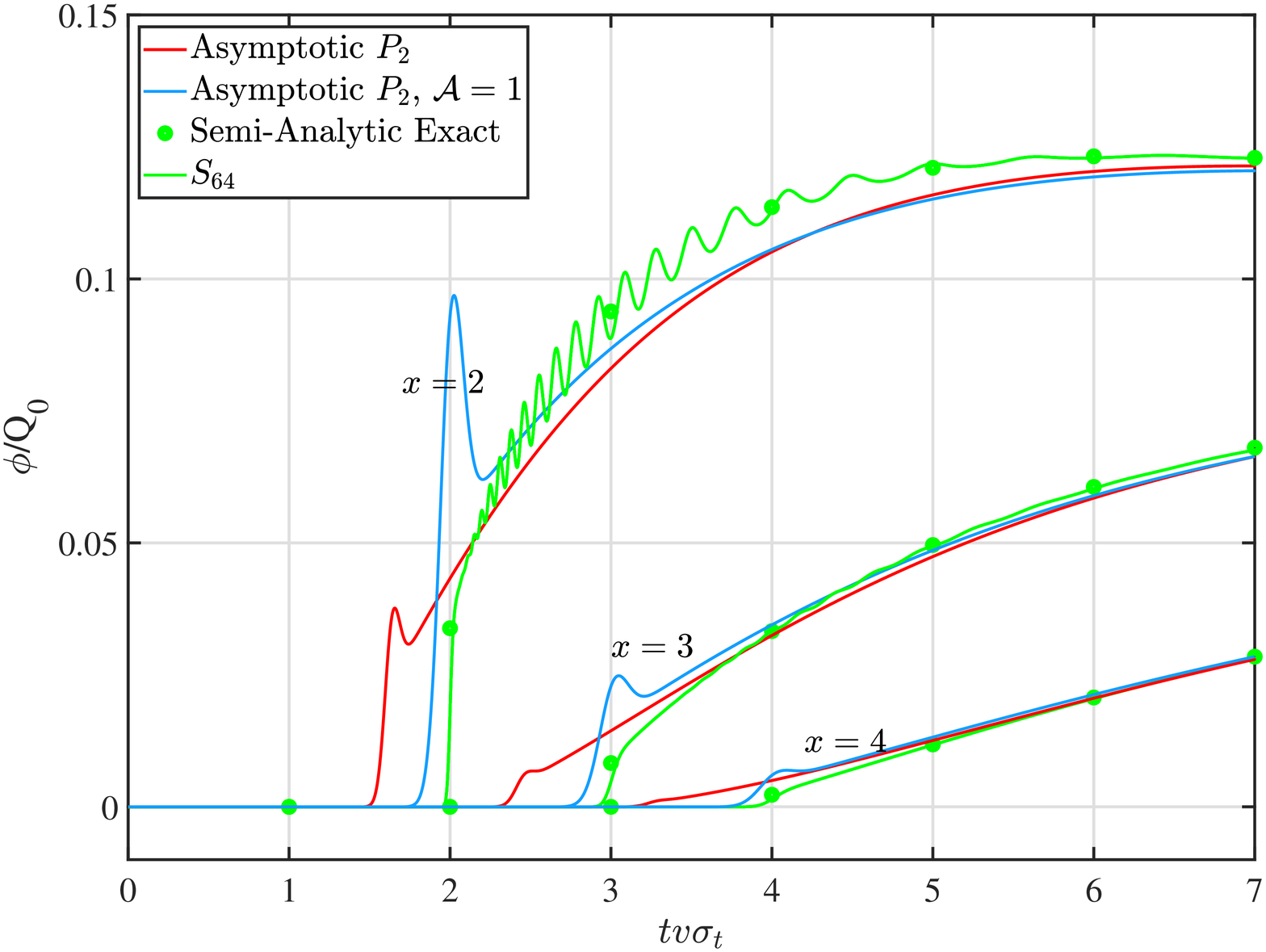}
\includegraphics*[width=7.5cm]{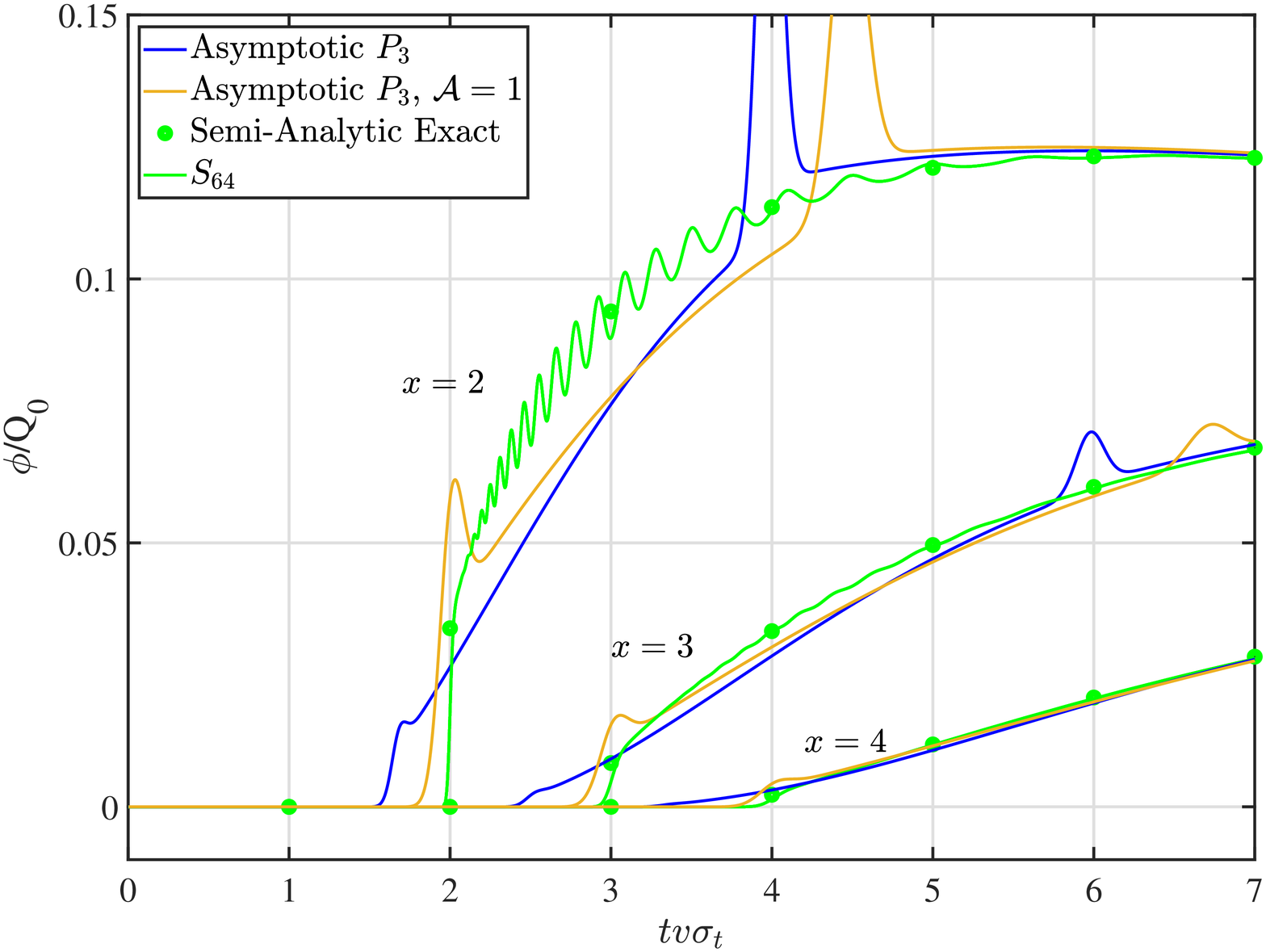}
\caption{A comparison between the asymptotic $P_N$ approximation to Pomraning's $P_{1/\B_N}$ approximation. The scalar flux of the different approximations as a function of the dimensionless time for the fully scattering case ($c=1$). The $P_{1/\B_N}$ approximation are in the light blue curves. The $P_2$ results is on the left figure, when the $P_3$ results are on the right.}
\label{fig:pomraning_ap}
\end{figure}
Another comparison is to the {\em ad hoc} Pomraning's proposal in~\cite{pomraning2005equations} (Eq. 3.147), which we call the $P_{1/\B_N}$ approximation. In this proposal, one take the {\em time-independent} closure, determining $\B_N$ exactly as in the time-dependent asymptotic $P_N$ approximation:
\begin{equation}\label{pom_B}
\frac{1}{\B_N}\equiv D_N = \frac{N+(N+1)\alpha_N}{(2N+1)}
\end{equation}
when $\alpha_N$ is determined by Eq.~\ref{alphapomraning} and setting $\A_N=1$. As a matter of fact, for purely scattering case $c=1$ this choice is the $N$'s expansion of the {\em ad hoc} $P_{1/3}$ approximation, which for $N=1$ is exactly Pomraning's proposal (for general medium other then $c=1$, is an expansion of the asymptotic $P_{1/3}$ approximation~\cite{heizler2012sp}). A similar {\em ad hoc} choice is in the works of Olson which proposal the exact $N$'s expansion of the {\em ad hoc} $P_{1/3}$ approximation~\cite{olson2012alternate, olson2019positivity} for any medium $c$.

In Fig.~\ref{fig:pomraning_ap} we compare the asymptotic time-dependent $P_N$ to Pomraning's $P_{1/\B_N}$ approximation. The $P_{1/\B_N}$ approximation are in the light blue curves, while the $P_2$ results are in the left figure and the $P_3$ results are on the right. Overall, the two approximations yield very close results. The only difference is in the tails, where the $P_{1/\B_N}$ approximation yields the exact particle velocity due to the $\A_N=1$ choice, and the asymptotic $P_N$ yields better results near the tail areas. This means that the time-dependent asymptotic $P_N$ gives the rigorous mathematical justification (due to asymptotic analysis derivation) for the purely {\em ad hoc} $P_{1/\B_N}$ approximation, due to the close value of $\A_N$ (and same value for $\B_N$).

\subsection{Substantial Absorbing Case}
In addition to the fully scattering case, we have tested the new approximation also for $c=0.5$ in order to demonstrate the behavior in an absorbing media. This case is interesting although the exact solution attains the scaling form in Eq.~\ref{scale}, however, the approximations do not, and the time for yielding the central limit theorem (the diffusion limit) is larger, so the differences in the tails zones between the different approximation, increase.
\begin{figure}[htbp!]
\includegraphics[width=17cm,trim={0 0 5cm 0},clip]{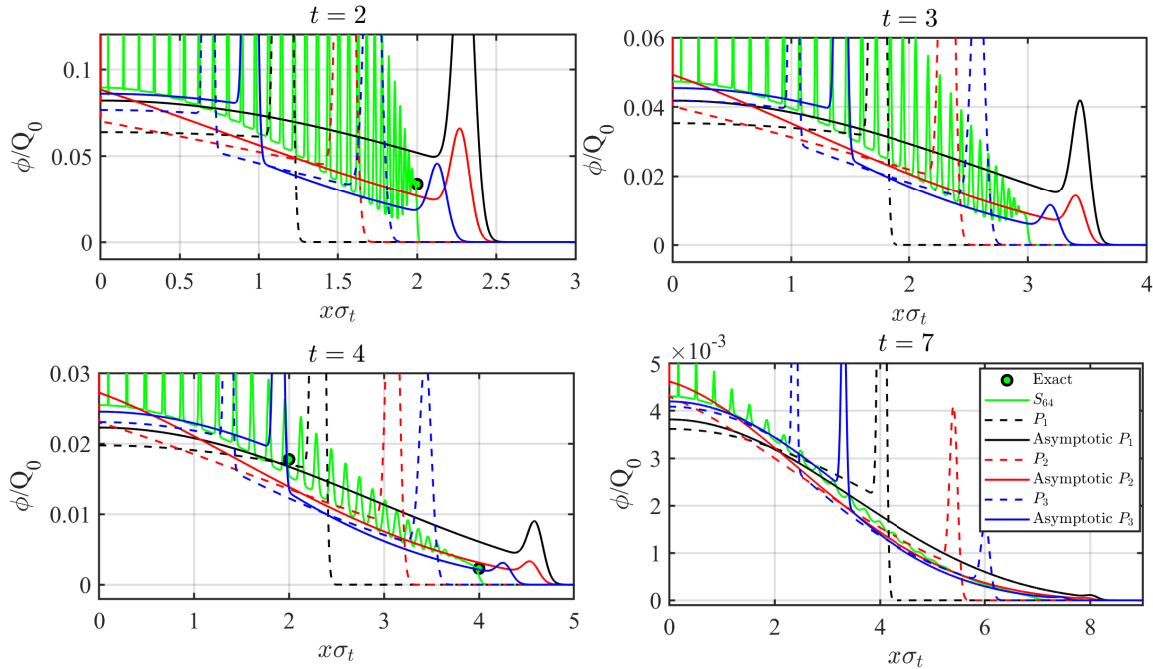}
\caption{The scalar flux of the different approximations and the exact benchmark as a function of dimensionless space for the substantial absorbing case ($c=0.5$).}
\label{fig:c05_t}
\end{figure}

\begin{figure}[htbp!]
\includegraphics*[width=7.5cm]{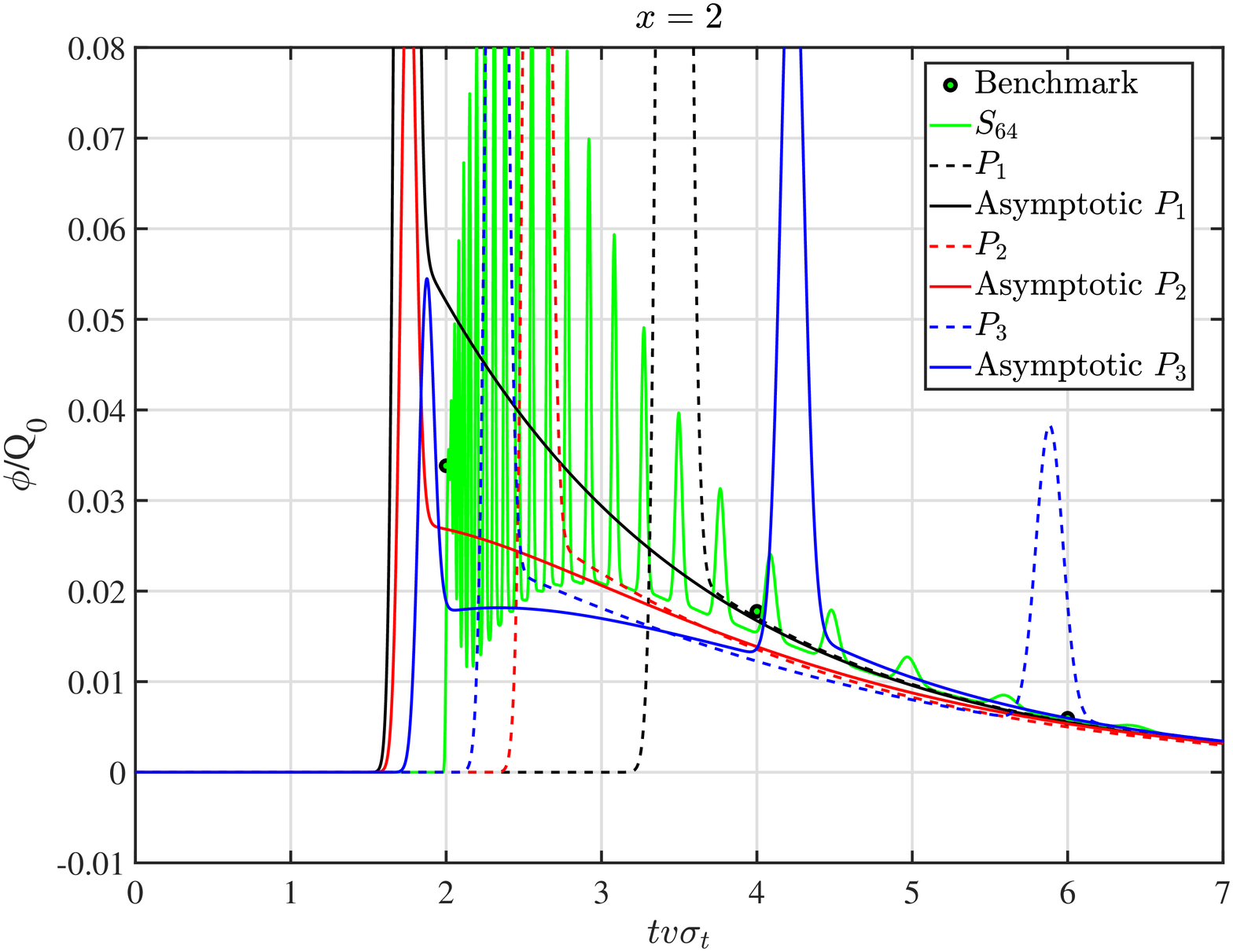}
\includegraphics*[width=7.5cm]{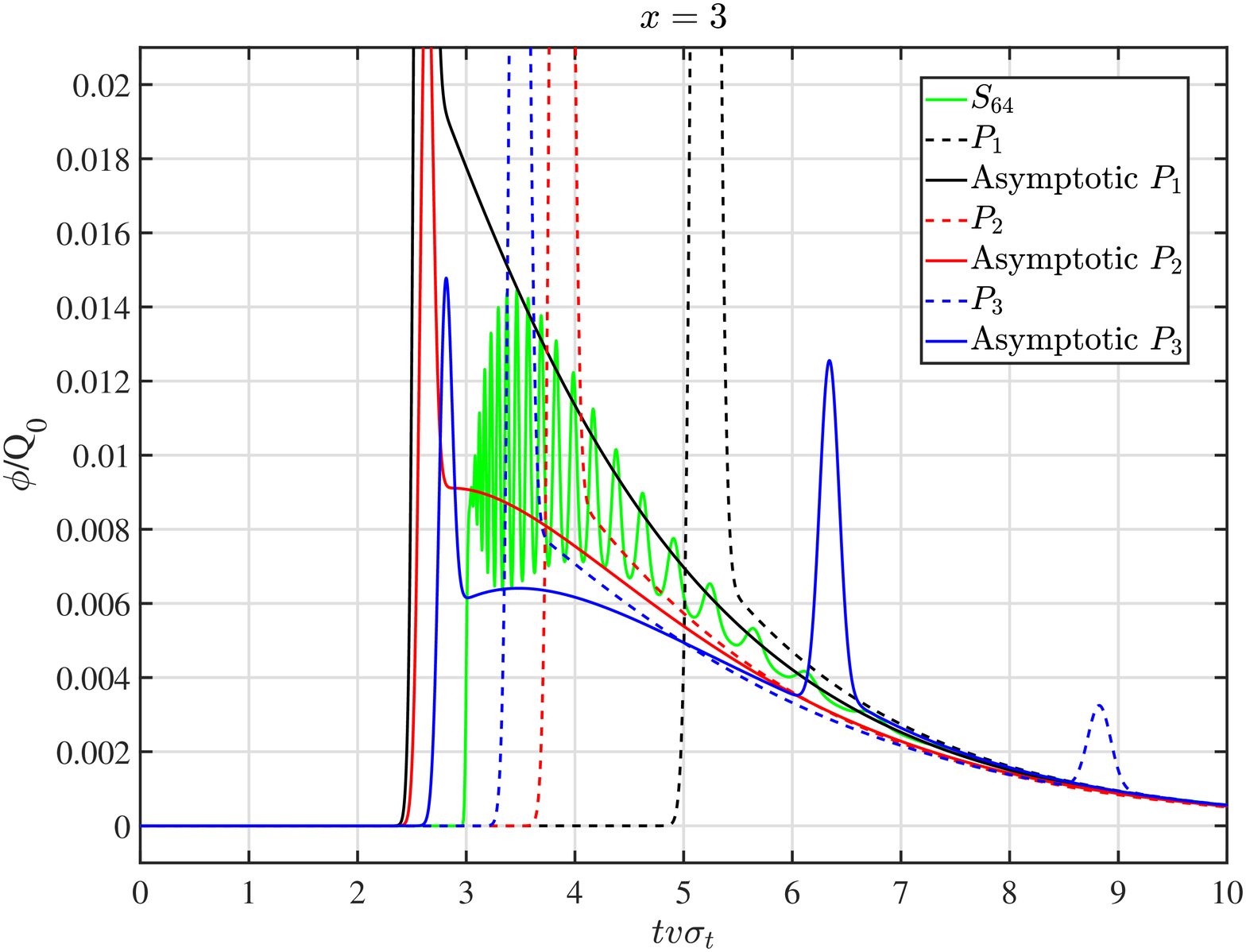}
\includegraphics*[width=7.5cm]{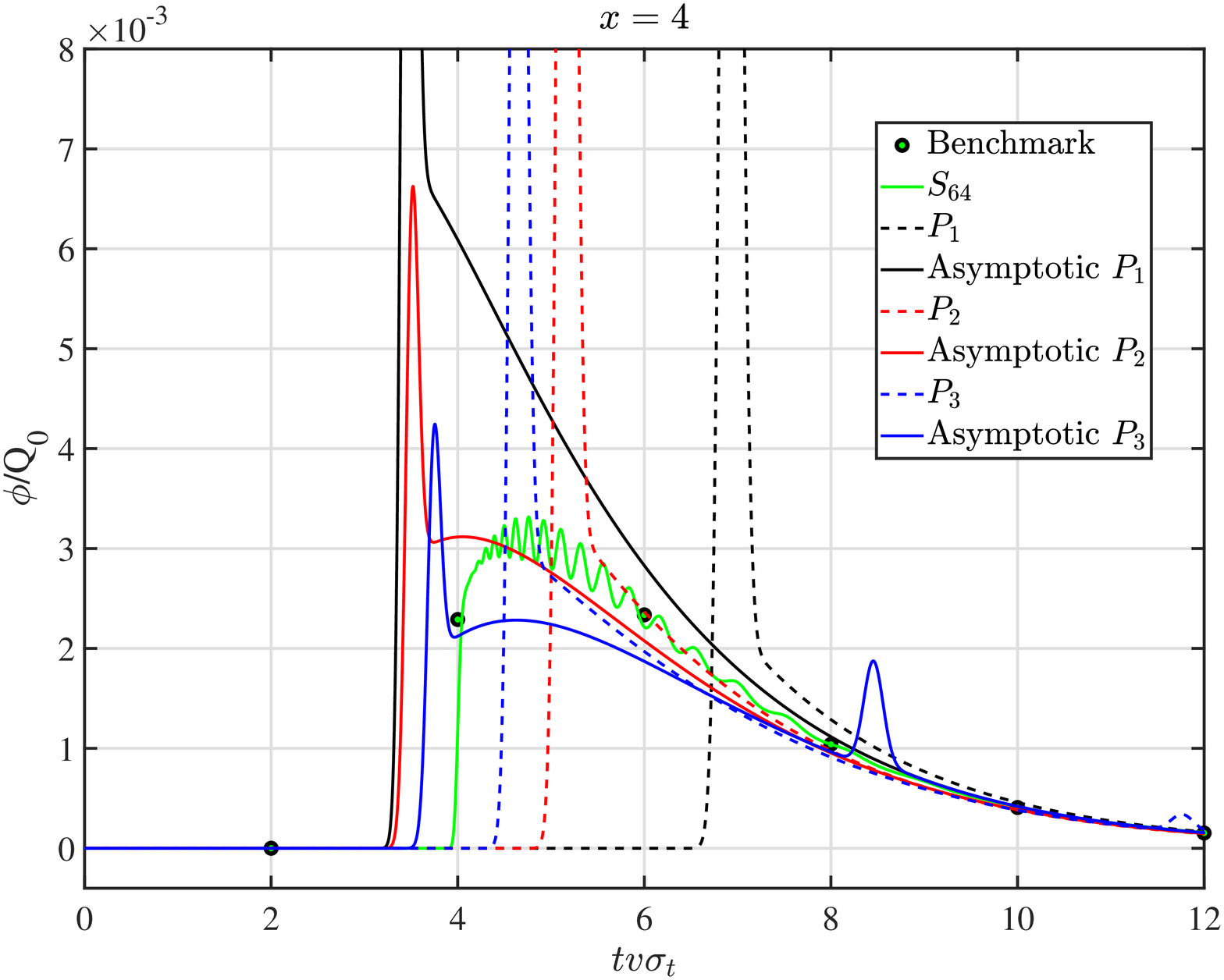} 
\caption{The scalar flux of the different approximations as a function of the dimensionless time for the substantial absorbing case ($c=0.5$).}
\label{fig:c05_x}
\end{figure}

In Fig.~\ref{fig:c05_t} the scalar flux is compared for different times ($t = 2,3,4$ and 7) while in Fig.~\ref{fig:c05_x} the scalar flux is compared for different positions ($x=2, 3$ and 4). From Figs.~\ref{fig:c05_t} and~\ref{fig:c05_x} we point out the differences between the two cases. In the substantial absorbing case the scalar flux drops to zero for every $x$ in sufficient $t$ unlike the fully scattering case. 
In both cases, the asymptotic time-dependent $P_N$ yields better results than its classic counter-part. 

\section{Concluding Remarks}
Since the exact Boltzmann equation is hard and time-consuming to solve, many approximations were introduced that are easier to solve. Here we focus on the spherical harmonics ($P_N$) approximation, when the exact equation is replaced by a finite closed set of equations for moment, with some closure condition, that is the heart of the approximation. 

We have derived a new approximation, namely {\em the time-dependent asymptotic $P_N$ approximation}, by providing a new closure equation. This approximation rests on two foundations: The time-dependent asymptotic $P_1$ from one hand~\cite{heizler2010asymptotic}, and the Pomraning's time-independent asymptotic $P_N$~\cite{pomraning1964generalized} on the other. In the same way as the asymptotic $P_1$ is the time-dependent generalization of the asymptotic diffusion equation, the proposed approximation is the time-dependent generalization of the Pomraning's time-independent asymptotic $P_N$ approximation.
We use the asymptotic solution of the exact Boltzmann equation, in both space and time, in infinite homogeneous medium, and derive a general $N$ closure equation, exploiting the Laplace domain, and the Case et al. spatial-asymptotic solution~\cite{case1953introduction,CaseZweifel1967}. Our closure equation contains two {\em linear} coefficients (and thus easy and stable to solve), $\A_N(c)$ and $\B_N(c)$, that are closed and known functions of the albedo $c$ (see Fig.~\ref{fig:AB}).

We have tested the time-dependent asymptotic $P_N$ approximation in an one-dimensional benchmark for the full Green function, and have shown that the new approximation yields better results than the classic $P_N$ for any given $N$. This test was done for both the case of fully scattering medium and for the case of the substantial absorbing medium. Even for $N=2$, it yields a very good approximate solution, even for the tails of the distribution. This new approximation yields also a better approximation as compared to other modified $P_N$ closures. As a matter of fact, the new time-dependent asymptotic $P_N$ provides the rigorous mathematical foundation, for the {\em ad hoc} approximate closures (the $P_{1/\B_N}$ approximations) that were found to be accurate in several famous problems.

\appendix
\setcounter{section}{-1}
\section{Explicit expressions for the dimensionless 
asymptotic diffusion coefficient $D_0$}\label{appendixA}
\renewcommand{\theequation}{A\arabic{equation}}
In this appendix we give the fitted explicit expressions for the dimensionless asymptotic diffusion coefficient $D_0(c)$ (Eq.~\ref{D0}) that depends on the solution to the transcendental equation for the eigenvalues $\varkappa_0(c)$ of the asymptotic transport equation, for infinite homogeneous medium (Eq.~\ref{kappa}).

For $c \ll 1$ which is highly absorbing media~\cite{case1953introduction}:
\begin{equation}\label{D0_small_c}
    D_0(c) \approx (1-c) \left[1+4e^{-\frac{2}{c}} + \frac{8(c+2)}{c}e^{-\frac{4}{c}} + \frac{12(8+4c+c^2)}{c^2}e^{-\frac{6}{c}} + ...\right]
\end{equation}
For $(1-c) \ll 1$ which is highly scattering media~\cite{case1953introduction}:
\begin{equation}\label{D0_big_c}
    D_0(c) \approx \frac{1}{3} \left[1 + \frac{4}{5}(1-c) + \frac{108}{175}(1-c)^2 +\frac{396}{875}(1-c)^3 + ...\right]
\end{equation}
In Winslow's paper~\cite{winslow1968} there is a suggestions for the value of $D_0(c)$ with a maximal error of 1\%:
\begin{equation}
        \label{D0_winslow}
        D_0(c) =\begin{cases}
         (1-c) & 
        c \leq 0.3 \\ 
        \frac{4}{\pi^2 c} \left(\frac{c-0.0854}{c+0.112}\right) & 
        c > 0.3
        \end{cases}
\end{equation}
Another approximation of $D_0$ from~\cite{dofer} with a maximal error of 0.3\%:
\begin{equation}
    D_0(c) = \begin{cases}
    0.80054- 0.523c & 
     \text{if} \, 0.59 \leq c \leq 0.61 \\
    \left(\frac{0.40528473}{1+c}\right) \frac{0.3267567 + c [0.1587312 - c(0.5665676 + c)]}{0.1326495 + c[0.03424169 + c(0.1774006 - c)]} &   \text{otherwise}
   \end{cases}
\end{equation}

These expression will be used inside Eq.~\ref{asymptotic_pn_ficklaw}, for solving the coefficients $\A_N(c)$ and $\B_N(c)$ for the given $N$, each one in a different valid regime of $c$. The results are given by Fig.~\ref{fig:AB}, when the curves were yield by a superposition of using the different fits.

\begin{acknowledgments}
We acknowledge the support of the PAZY Foundation under Grant \textnumero~61139927. SIH wishes to dedicate this work to Piero Ravetto, who accompany my work of transport theory over the last decade, and actually gave the motivation to this work, using a general $N$.
\end{acknowledgments}
\setlength{\baselineskip}{12pt}
\bibliographystyle{unsrtnat} 
\bibliography{bibliography} 

\begin{thebibliography}{46}
\providecommand{\natexlab}[1]{#1}
\providecommand{\url}[1]{\texttt{#1}}
\expandafter\ifx\csname urlstyle\endcsname\relax
  \providecommand{\doi}[1]{doi: #1}\else
  \providecommand{\doi}{doi: \begingroup \urlstyle{rm}\Url}\fi

\bibitem[Duderstadt and Martin(1979)]{Duderstadt}
James~J. Duderstadt and William~R. Martin.
\newblock \emph{Transport Theory}.
\newblock John Wiley \& Sons, 1979.

\bibitem[Epperlein(1994)]{epperlein1994}
E.M. Epperlein.
\newblock Fokker-planck modeling of electron transport in laser-produced
  plasmas.
\newblock \emph{Laser and Particle Beams}, 12\penalty0 (2):\penalty0 257--272,
  1994.

\bibitem[Castor(2004)]{castor2004}
John~I. Castor.
\newblock \emph{Radiation Hydrodynamics}.
\newblock Cambridge University Press, 2004.

\bibitem[Pomraning(1973)]{pomraning2005equations}
Gerald~C. Pomraning.
\newblock \emph{The equations of radiation hydrodynamics}.
\newblock Pergamon Press, 1973.

\bibitem[Lindl et~al.(2004)Lindl, Amendt, Berger, Glendinning, Glenzer, Haan,
  Kauffman, Landen, and Suter]{lindl2004}
John~D. Lindl, Peter Amendt, Richard~L. Berger, S.~Gail Glendinning,
  Siegfried~H. Glenzer, Steven~W. Haan, Robert~L. Kauffman, Otto~L. Landen, and
  Laurence~J. Suter.
\newblock The physics basis for ignition using indirect-drive targets on the
  national ignition facility.
\newblock \emph{Physics of Plasmas}, 11\penalty0 (2):\penalty0 339--491, 2004.

\bibitem[Rosen(1996)]{rosen1996}
Mordecai~D. Rosen.
\newblock The science applications of the high-energy density plasmas created
  on the nova laser.
\newblock \emph{Physics of Plasmas}, 3\penalty0 (5):\penalty0 1803--1812, 1996.

\bibitem[Cohen et~al.(2020)Cohen, Malamud, and Heizler]{prr}
Avner~P. Cohen, Guy Malamud, and Shay~I. Heizler.
\newblock Key to understanding supersonic radiative marshak waves using simple
  models and advanced simulations.
\newblock \emph{Phys. Rev. Research}, 2:\penalty0 023007, 2020.

\bibitem[Case and Zweifel(1967)]{CaseZweifel1967}
K.M. Case and P.F. Zweifel.
\newblock \emph{Linear Transport Theory}.
\newblock Addison-Wesley Publishing Company, 1967.

\bibitem[Bell and Glasstone(1970)]{BellGlasstone}
G.~I. Bell and S.~Glasstone.
\newblock \emph{Nuclear Reactor Theory}.
\newblock Van Nostrand Reinhold Co., 10 1970.

\bibitem[Weiss(2002)]{weiss2002}
George~H. Weiss.
\newblock Some applications of persistent random walks and the telegrapher's
  equation.
\newblock \emph{Physica A: Statistical Mechanics and its Applications},
  311\penalty0 (3):\penalty0 381 -- 410, 2002.

\bibitem[J. and K.(2017)]{masoliver}
Masoliver J. and Lindenberg K.
\newblock Continuous time persistent random walk: a review and some
  generalizations.
\newblock \emph{Eur. Phys. J. B}, 90:\penalty0 107, 2017.

\bibitem[Pawula(1967)]{pawula1967approximation}
R.F. Pawula.
\newblock Approximation of the linear boltzmann equation by the fokker-plank
  equation.
\newblock \emph{Phys. Rev.}, 162:\penalty0 196, 1967.

\bibitem[Carlson and Bell(1958)]{sn}
B.G. Carlson and G.I. Bell.
\newblock \emph{Solution of the Transport Equation by the $S_n$ Method}.
\newblock A/CONF.15/P/2386, Los Alamos Scientific Laboratory, NM, 1958.

\bibitem[Fleck and Cummings(1971)]{imc}
J.A. Fleck and J.D. Cummings.
\newblock An implicit monte carlo scheme for calculating time and frequency
  dependent nonlinear radiation transport.
\newblock \emph{Journal of Computational Physics}, 8\penalty0 (3):\penalty0 313
  -- 342, 1971.

\bibitem[Sch{\"a}fer et~al.(2011)Sch{\"a}fer, Frank, and
  Levermore]{levermore2011}
Matthias Sch{\"a}fer, Martin Frank, and C.~David Levermore.
\newblock Diffusive corrections to $p_n$ approximations.
\newblock \emph{Multiscale Modeling \& Simulation}, 9\penalty0 (1):\penalty0
  1--28, 2011.

\bibitem[Zheng and McClarren(2016)]{mcclarren2016}
Weixiong Zheng and Ryan~G. McClarren.
\newblock Moment closures based on minimizing the residual of the pn angular
  expansion in radiation transport.
\newblock \emph{Journal of Computational Physics}, 314:\penalty0 682 -- 699,
  2016.

\bibitem[Case et~al.(1953)Case, Hoffmann, and Placzek]{case1953introduction}
K.M. Case, F.De Hoffmann, and G.~Placzek.
\newblock \emph{Introduction to the theory of neutron diffusion}, volume~1.
\newblock Los Alamos Scientific Laboratory, NM, 1953.

\bibitem[Bengston(1958)]{bengston1958}
J.~Bengston.
\newblock \emph{The asymptotic time-dependent transport equation}.
\newblock UCRL-5209, University of California Radiation Laboratory, Berkeley
  CA, 1958.

\bibitem[Heizler(2010)]{heizler2010asymptotic}
Shay~I Heizler.
\newblock Asymptotic telegrapher's equation ($p_1$) approximation for the
  transport equation.
\newblock \emph{Nuclear science and engineering}, 166\penalty0 (1):\penalty0
  17--35, 2010.

\bibitem[Heizler(2012)]{heizler2012}
Shay~I. Heizler.
\newblock The asymptotic telegrapher's equation ($p_1$) approximation for
  time--dependent, thermal radiative transfer.
\newblock \emph{Transport Theory and Statistical Physics}, 41\penalty0
  (3-4):\penalty0 175--199, 2012.

\bibitem[Heizler and Ravetto(2012)]{heizler2012sp}
Shay~I Heizler and Piero Ravetto.
\newblock $sp_2$--asymptotic $p_1$ equivalence.
\newblock \emph{Transport Theory and Statistical Physics}, 41\penalty0
  (3-4):\penalty0 304--324, 2012.

\bibitem[Cohen et~al.(2018)Cohen, Perry, and Heizler]{cohen2018discontinuous}
Avner~P. Cohen, Roy Perry, and Shay~I. Heizler.
\newblock The discontinuous asymptotic telegrapher's equation ($p_1$)
  approximation.
\newblock \emph{Nuclear Science and Engineering}, 192\penalty0 (2):\penalty0
  189--207, 2018.

\bibitem[Cohen and Heizler(2018)]{cohen2019discontinuous}
Avner~P. Cohen and Shay~I. Heizler.
\newblock Modeling of supersonic radiative marshak waves using simple models
  and advanced simulations.
\newblock \emph{Journal of Computational and Theoretical Transport},
  47\penalty0 (4-6):\penalty0 378--399, 2018.

\bibitem[Pomraning(1964)]{pomraning1964generalized}
G.C. Pomraning.
\newblock A generalized pn approximation for neutron transport problems.
\newblock \emph{Nukleonik}, 6, 1964.

\bibitem[Ganapol(1999)]{ganapol1999}
B.D. Ganapol.
\newblock \emph{Homogeneous Infinite Media Time- Dependent Analytic Benchmarks
  for X-TM Transport Methods Development}.
\newblock Technical Report, Los Alamos National Laboratory, NM, 1999.

\bibitem[Olson and Henderson(2004)]{olson2004numerical}
K.R. Olson and D.L. Henderson.
\newblock Numerical benchmark solutions for time-dependent neutral particle
  transport in one-dimensional homogeneous media using integral transport.
\newblock \emph{Annals of Nuclear Energy}, 31\penalty0 (13):\penalty0
  1495--1537, 2004.

\bibitem[Davison and Sykes(1957)]{davison1957neutron}
Boris Davison and John~Bradbury Sykes.
\newblock \emph{Neutron transport theory}.
\newblock Oxford University Press, 1957.

\bibitem[snd P.~Ravetto(1980)]{CoppaRavetto1980}
G.~Coppa snd P.~Ravetto.
\newblock \emph{Il Metodo Dell'espansione in Autostati per la Soluzione di
  Problemi Dinamici di Transporto per i Neutroni Nell'approssimazione $P_N$,
  (in Italian)}.
\newblock Politecnico di Torino, PT-IN-FR-116, 1980.

\bibitem[Rulko and Larsen(1993)]{RulkoLarsen1993}
Robert~P. Rulko and Edward~W. Larsen.
\newblock Variational derivation and numerical analysis of p2 theory in planar
  geometry.
\newblock \emph{Nuclear Science and Engineering}, 114\penalty0 (4):\penalty0
  271--285, 1993.

\bibitem[U.~Shin and Morel(1993)]{ShinMorel1993}
W.F.~Miller U.~Shin and J.E. Morel.
\newblock \emph{Asymptotic derivation of the modified time-dependent $SP_2$
  equations and numerical calculations}.
\newblock LA-UR-93-2301, Los Alamos Scientific Laboratory, NM, 1993.

\bibitem[Pomraning et~al.(1994)Pomraning, Rulko, and Su]{PomraningRulkoSu1994}
G.~C. Pomraning, Robert Rulko, and Bingjing Su.
\newblock Diffusion theory with discontinuities.
\newblock \emph{Nuclear Science and Engineering}, 118\penalty0 (1):\penalty0
  1--23, 1994.

\bibitem[Rulko(1995)]{Rulko1995}
Robert~P. Rulko.
\newblock Variational derivation of multigroup $p_2$ equations and boundary
  conditions in planar geometry.
\newblock \emph{Nuclear Science and Engineering}, 121\penalty0 (3):\penalty0
  371--392, 1995.

\bibitem[Dawson(1964)]{dawson}
Charles Dawson.
\newblock \emph{Modified $P_2$ Approximations to the Neutron Transport
  Equation}.
\newblock DTMB-1814, U.S. Department of the Navy, 1964.

\bibitem[Pomraning(1965)]{pomraning1965asymptotically}
G.C. Pomraning.
\newblock An asymptotically correct approximation to the multidimensional
  transport equation.
\newblock \emph{Nuclear Science and Engineering}, 22\penalty0 (3):\penalty0
  328--338, 1965.

\bibitem[Winslow(1968)]{winslow1968}
Alan~M. Winslow.
\newblock Extensions of asymptotic neutron diffusion theory.
\newblock \emph{Nuclear Science and Engineering}, 32\penalty0 (1):\penalty0
  101--110, 1968.

\bibitem[Zimmerman(1979)]{zimmerman1979}
G.B. Zimmerman.
\newblock \emph{Differencing Asymptotic Diffusion Theory}.
\newblock UCRL-82792, Lawrence Livermore Laboratory, University of CA
  Livermore, 1979.

\bibitem[Ganapol and Pomraning(1996)]{GanapolPomraning1996}
B.D. Ganapol and G.~C. Pomraning.
\newblock The two-region milne problem.
\newblock \emph{Nuclear Science and Engineering}, 123\penalty0 (1):\penalty0
  110--120, 1996.

\bibitem[Huang and Lewis(1972)]{huang1972asymptotic}
Song~Teh Huang and Elmer~E. Lewis.
\newblock Asymptotic pn and double pn approximations.
\newblock \emph{Journal of Nuclear Energy}, 26\penalty0 (5):\penalty0 231--236,
  1972.

\bibitem[Morel et~al.(2013)Morel, Ragusa, Adams, and
  Kanschat]{morel2013asymptotic}
Jim~E. Morel, Jean~C. Ragusa, Marvin~L. Adams, and Guido Kanschat.
\newblock Asymptotic $p_n$-equivalent $s_{N+1}$ equations.
\newblock \emph{Transport Theory and Statistical Physics}, 42\penalty0
  (1):\penalty0 3--20, 2013.

\bibitem[Olson et~al.(2000)Olson, Auer, and Hall]{p13}
Gordon~L. Olson, Lawrence~H. Auer, and Michael~L. Hall.
\newblock Diffusion, $p_1$, and other approximate forms of radiation transport.
\newblock \emph{Journal of Quantitative Spectroscopy and Radiative Transfer},
  64\penalty0 (6):\penalty0 619 -- 634, 2000.

\bibitem[Olson(2012)]{olson2012alternate}
Gordon~L. Olson.
\newblock Alternate closures for radiation transport using legendre polynomials
  in 1d and spherical harmonics in 2d.
\newblock \emph{Journal of Computational Physics}, 231\penalty0 (7):\penalty0
  2786--2793, 2012.

\bibitem[Olson(2019)]{olson2019positivity}
Gordon~L. Olson.
\newblock Positivity enhancements to the pn transport equations.
\newblock \emph{Journal of Computational and Theoretical Transport},
  48\penalty0 (5):\penalty0 159--179, 2019.

\bibitem[M.~Frank and Yasuda(2007)]{spn}
E.W.~Larsen M.~Frank, A.~Klar and S.~Yasuda.
\newblock Time-dependent simplified $p_n$ approximation to the equations of
  radiative transfer.
\newblock \emph{J. Comp. Phys.}, 226:\penalty0 2289--2305, 2007.

\bibitem[Jackson(1943)]{legendre}
Dunham Jackson.
\newblock Legendre functions of the second kind and related functions.
\newblock \emph{The American Mathematical Monthly}, 50\penalty0 (5):\penalty0
  291--302, 1943.

\bibitem[Ku{\v s}{\v c}er and Zweifel(1965)]{Kuscer1965}
I.~Ku{\v s}{\v c}er and P.~F. Zweifel.
\newblock Time--dependent one--speed albedo problem for a semi-infinite medium.
\newblock \emph{Journal of Mathematical Physics}, 6\penalty0 (7):\penalty0
  1125--1130, 1965.

\bibitem[Ofer()]{dofer}
Dror Ofer.
\newblock Private Communication.

\end{thebibliography}

\end{document}